\documentclass{jfm}
\usepackage{graphicx}
\usepackage{epstopdf, epsfig}
\usepackage{amsmath,amssymb,amsfonts}
\usepackage{float}
\usepackage{subfig,xcolor}

\shorttitle{Thin-gap averaging of variable-viscosity flows}
\shortauthor{Pillai, Picardo and Narayanan}

\title{Thin-gap averaging of variable-viscosity flows: application to thermoviscous fingering}

\author{Dipin S. Pillai\aff{1}\corresp{\email{dipinsp@iitk.ac.in}},
 Jason R. Picardo\aff{2} \corresp{Associate, International Centre for Theoretical Sciences, TIFR, Bangalore, India.}
 \and R. Narayanan\aff{3}
  }

\affiliation{\aff{1}Department of Chemical Engineering, Indian Institute of Technology Kanpur, Uttar Pradesh, 208016, India\\
\aff{2}Department of Chemical Engineering, Indian Institute of Technology Bombay, Mumbai, 400 076, India\\
\aff{3}Department of Chemical Engineering, University of Florida,
Gainesville, FL 32611, USA}

\begin{document}

\maketitle

\begin{abstract}
A consistent averaging technique, using the weighted residual integral boundary layer (WRIBL) method, is presented for flow through a thin-gap geometry wherein the fluid's viscosity varies across the gap. In such situations, the flow has a non-parabolic cross-gap velocity profile---an effect that is ignored by Darcy models conventionally used for such Hele-Shaw flows. The WRIBL technique systematically accounts for the cross-gap variation of viscosity and yields reduced-order equations for the gap-averaged fluid flow rate. As a test case, we consider a fluid with a temperature-dependent viscosity and analyse the previously-studied problem of thermoviscous fingering: a hot fluid flowing through a Hele-Shaw geometry with cold walls spontaneously forms channels of low-viscosity, hot fluid, separated by regions of high-viscosity, cold fluid.  The temperature of the cold walls is assumed either to be constant, a scenario that mimics the upward flow of magma through fissures in the Earth's crust, or to vary linearly along the direction of flow. In both cases, the predictions of the WRIBL model, regarding the multiplicity of uniform steady flow states and their linear stability, are compared with that of the hitherto used, \textit{ad hoc}, Darcy model (Helfrich, J. Fluid Mech., vol. 305, 1995, pp. 219-238), as well as with calculations of the full three-dimensional governing equations (Wylie \& Lister, J. Fluid Mech., vol. 305, 1995, pp. 239-261).
Though the results are qualitatively similar, the WRIBL model is found to be much more accurate than the Darcy model. The averaging method is presented in a general manner to facilitate its application to other physical situations where, for example, the viscosity depends on solute/particle concentration in a solution/suspension.
\end{abstract}


\section{Introduction}
Several physico-chemical problems offer instances of transport phenomena taking place in confined geometries, such as a Hele-Shaw cell, wherein one of the spatial dimensions is much smaller than the other two. In such cases, a convenient and often-used strategy is to employ a thin-gap averaged model based on a parabolic cross-gap velocity profile to describe the underlying physics. This yields a simplified Darcy-like averaged model, which then serves as an efficient tool for understanding the behavior of such systems \citep{Homsy1987}. However, one often encounters instances wherein the cross-gap velocity profile may significantly deviate from the Poiseuille parabolic flow. As an example, this is true when the flow is coupled to the transport of a scalar, such as temperature. This scenario occurs in the case of basaltic magma flow through fissures, where cross-gap variations in temperature result in a non-uniform viscosity across the thin gap \citep{Wylie1995}. Even in such cases, though, \textit{ad hoc} thin-gap averaging based on a parabolic velocity profile is often employed \citep{Pearson1973, Whitehead1991, Bercovici1994, Helfrich1995}. Other situations where this issue arises include Hele-Shaw flows of suspensions of colloids, non-Brownian particles, or microswimmers with a number concentration that varies across the thin gap~\citep{Lyon1998,Marmet2017,Laxmi2020}, thus resulting in a non-uniform effective viscosity and a non-parabolic flow profile. 

There are of course special cases where no cross-gap variations occur, for example, when adiabatic or no-flux conditions are imposed on the sidewalls \citep{Pritchard2009}. However, in general, this is not true, as non-uniformities can arise due to transport across the boundaries, as when the fluid loses heat through the side walls \citep{Helfrich1995,Nagatsu2009}, or due to gradient-producing physical effects, such as shear-induced migration in suspensions~\citep{Acrivos1987,Nott1994}, wall effects~\citep{Marmet2017}, and phoretic motion~\citep{Buongiorno2005,choudhary2020}. The inherent inconsistency in using Darcy models in such situations has been recognized previously; however, in the past, one has had to either embrace the complexity of the full three-dimensional equations \citep{Wylie1995} or treat the problem in a linearized limit \citep{Viola2017}. In this work, we address this issue by developing a low-dimensional modelling strategy that takes advantage of the thin-gap geometry while systematically accounting for cross-gap variations in viscosity and the consequent non-parabolic velocity profile.

Given the wide range of possible applications, we first present the averaging method in a general setting in \S\ref{sec:mathmodel} for an arbitrary cross-gap variation of viscosity. Our approach is based on  the weighted residual integral boundary layer (WRIBL) method, which was developed to model the inertial dynamics of thin liquid films \citep{RQ2002}, and has been subsequently used in a variety of free-surface and two-phase flow problems \citep{Kalliadasis,Trev2004,Oron2008,Dietze2013}, including cases where the fluid is non-Newtonian~\citep{RQ2012,Pradas2014}. This averaging method has also been used to incorporate inertial corrections into the Darcy model for constant-viscosity Hele-Shaw flows~\citep{RQ2001}. The first step of the WRIBL procedure is to obtain simplified long-wave partial differential equations from the full incompressible Navier-Stokes equations, based on the disparity of cross-gap and in-plane length scales. These equations are then subsequently averaged across the thin gap using appropriate weight functions. While several weighting strategies are possible, the Galerkin weighting method was shown by \cite{RQ2002} to be the most efficient and we employ the same. The key step in the method is choosing the weight functions so that an asymptotically consistent, closed system of averaged equations are obtained; it is here that the effect of a non-uniform viscosity is incorporated.

\begin{figure}
  \centerline{\includegraphics[width=10cm]{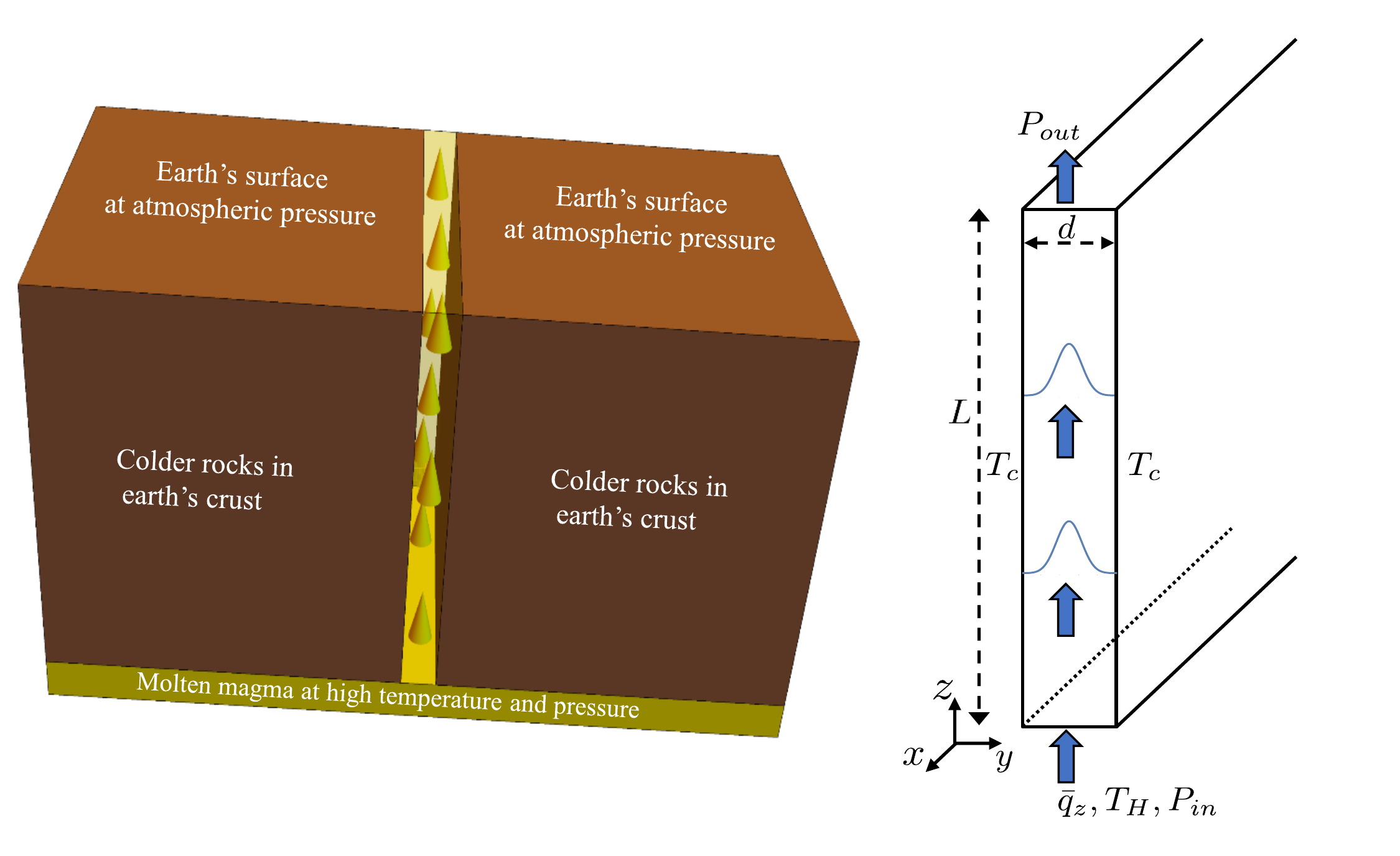}}
  \caption{Schematic of the Hele-Shaw flow system (right) that models magma flowing upward through a fissure in the earth's crust (left). The thin-gap dimension of the fissure is given by $d$, and the total length is $L$ (with a span-wise dimension of similar magnitude) such that $d\ll L$. At the bottom inlet, the pressure is $P_{in}$ while the temperature is $T_H$. The pressure at the outlet is $P_{out}$. We consider a fixed pressure drop $P_{out}-P_{in}$ that drives a flow rate $q_z$ in the uniform base flow state. The surrounding rocks are at a colder temperature, which is assumed to have a constant value $T_C$. In a second version of the problem we allow the wall temperature to vary as $T_C+(T_H-T_C) (z/L)$.}
\label{fig:schematic}
\end{figure}

After developing the general averaged flow model, we then apply it to the problem of thermoviscous fingering in a Hele-Shaw geometry~\citep{Helfrich1995,Wylie1995}. This problem serves as an example to elucidate the significance of incorporating cross-gap variations during the averaging procedure. Figure \ref{fig:schematic} shows a schematic of the system, in which a pressure drop drives the flow of a hot fluid through a narrow slot, whose side-walls are held at a fixed lower temperature. This mimics the upward flow of pressurized hot magma through a narrow fissure whose walls, composed of relatively cold rock, cool the magma as it flows \citep{Wylie1999}. In a simplified model the magma is taken to be a Newtonian fluid with a viscosity that decreases sharply as it cools \citep{Huppert2002,Balm2001}---a property that nonlinearly couples flow and heat transport. For a sufficiently strong temperature-dependence of viscosity, the laterally-uniform steady flow is linearly unstable over a range of flow rates, giving rise to laterally alternating bands of low-viscosity (fast-moving) and high viscosity (slow-moving) magma~\citep{Helfrich1995,Wylie1995}.

This problem serves as a useful testing ground for the proposed averaging method because the Dirichlet boundary conditions applied at the cold walls cause the temperature to vary across the thin gap, which in turn produces a non-trivial variation in viscosity. We consider two cases, distinguished by the temperature of the walls which is assumed to be either constant or linearly decreasing along the direction of flow. While the latter scenario has no geological basis, it serves to emphasize the importance of cross-gap viscosity variations, possibly being relevant to industrial processes such as injection moulding and glass manufacturing in which the walls are actively cooled. In \S~\ref{sec:TVFmathmodel}, the WRIBL method is used to average the advection-diffusion equation for temperature, which is then coupled to the averaged flow equations of \S\ref{sec:mathmodel} to yield the final coupled WRIBL model. 

The Darcy model of \citet{Helfrich1995} is presented in \S~\ref{sec:Darcy}, where we discuss the ad hoc assumptions that must be made to obtain it from the WRIBL model.

In \S\ref{sec:results}, we compare the predictions of the coupled WRIBL model with that of the Darcy model, as well as with calculations of the full three dimensional (3D) equations. We focus on the multiplicity of the uniform steady base state and on its linear stability. The WRIBL model is found to be much more accurate than the Darcy model, with the latter significantly over-predicting the range of multiplicity and instability. However, the key qualitative features of the 3D results are captured by both averaged models. Therefore, while the Darcy model is found to be adequate for qualitative analysis, incorporating cross-gap variations of viscosity during the averaging procedure is important for making quantitative estimates.



\section{WRIBL averaging for variable-viscosity flows}\label{sec:mathmodel}

In this section, we present the depth-averaging strategy for the flow of a fluid with variable viscosity in a thin-gap geometry. For the sake of generality, we consider an arbitrary cross-gap variation of viscosity. The averaged model can be then used in different physical settings by coupling it to an additional conservation equation for the field variable on which the viscosity depends, as is done for temperature-dependent viscosity in \S\ref{sec:TVFmathmodel}. 

A Cartesian co-ordinate system is adopted, with $(u,v,w)$ being the components of the velocity vector ($\vec{v}$) in the ($x,y,z$) directions (figure~\ref{fig:schematic}). The depth ($d$) of the channel in the wall-bounded $y$ direction is much smaller than its length ($L$) along the primary flow direction ($z$), as well as its width in the lateral direction ($x$). We assume that all field variables, including the velocity and viscosity, vary much more slowly in the  $x$ and $z$ directions than in the thin-gap $y$ direction. Exploiting this disparity in length scales, a nonlinear reduced-order model, based on the WRIBL method, is developed for the in-plane, gap-averaged, flow rates $q_z(z,x,t)$ and $q_x(z,x,t)$. 

The flow is governed by the equation of continuity, and the Navier-Stokes equation with variable viscosity, $\mu(y;z,x,t)$, given below in dimensional form:
\begin{equation}\label{eq:gengeqn}
\nabla^*\cdot \vec{v}^*=0\qquad \text{and}\qquad \rho\left(\partial_{t^*}\vec{v}^*+ \vec{v}^* \cdot\nabla^* \vec{v}^*\right)=-\nabla^* p^* +\nabla^*\cdot\textbf{S}^*,
\end{equation}
where, $\rho$ is the fluid density, $\textbf{S}^*=\mu (\nabla^*\vec{v}^*+\nabla^*\vec{v}^{*T})$ is the deviatoric part of the Cauchy stress tensor, and $p^*$ is the isotropic fluid pressure. The asterisk ($^*$) denotes dimensional variables. 

At the walls, we apply no-slip boundary conditions: 
\begin{equation}
\vec{v}^*=0, \;\;{\rm at} \;\; y^*=0 \;\; {\rm and} \;\; y^*=d.
\end{equation}

The following scaling is used to non-dimensionalize the governing equations:
\begin{equation}
\begin{split}
&\ell_z=\ell_x=L,\quad\ell_y=d,\quad \mathcal{U}_z=\mathcal{U}_x=U,\quad\mathcal{U}_y=\varepsilon U, \quad t_c=d/ \varepsilon U,\\ &\mu_c=\mu_0,\,\,p_c=\mu_0 U/ d,
\end{split}
\label{eq:scaling}
\end{equation}
where the subscript $c$ denotes a characteristic scale, and $\epsilon$ is the long-wave parameter defined as $\varepsilon\equiv d/L\ll1$. Also $\mu_0$ is a characteristic value of viscosity for the system, for example an extremal or averaged value, and $U$ is a characteristic velocity. The scaled governing equations and boundary conditions, retaining terms only up to $\mathcal{O}(\varepsilon)$ and ignoring the higher order terms, reduce to
\begin{subequations}\label{eq:genlw}
\begin{equation}\label{eq:genlwcont}
\partial_x u+\partial_y v+\partial_z w=0,
\end{equation}
\begin{equation}\label{eq:genlwfl}
\partial_y p=0,\quad	-\varepsilon\partial_z p +\partial_y \left(\mu\partial_yw\right)=0, \quad -\varepsilon\partial_x p +\partial_y \left(\mu\partial_yu\right)=0,
\end{equation}
\begin{equation}\label{eq:genswbc}
u|_{y=0}=v|_{y=0}=w|_{y=0}=u|_{y=1}=v|_{y=1}=w|_{y=1}=0.
\end{equation}
\end{subequations}
In obtaining the above equations, the Reynolds number $\Rey = \rho d U/\mu_0$ is assumed to be small enough for inertial terms to be disregarded. Also, we assume that $\partial_y \mu \sim O(1)$ while  $\partial_x \mu \sim O(\varepsilon)$ and  $\partial_z \mu \sim O(\varepsilon)$, which requires the field variable (e.g., temperature) on which the viscosity depends to also vary slowly in the $x$ and $z$ directions. Typically, this field variable will be governed by an additional conservation equation which must also admit this separation of cross-gap and in-plane length scales.

From equation~\eqref{eq:genlwfl} we see that neglecting variations of $\mu$ in the cross-gap ($y$) direction, as is done in Darcy models, introduces $O(1)$ errors via $\partial_y \mu \partial_y u$ and $\partial_y \mu \partial_y w$ terms. Here, we consistently incorporate all leading order terms using
the weighted residual integral boundary layer (WRIBL) technique, to obtain the reduced-order nonlinear model up to $O(\varepsilon)$. We follow the procedure described in \citet{Dietze2013}, and begin by decomposing the velocity field into $\mathcal{O}(1)$ and $\mathcal{O}(\varepsilon)$ parts:
\begin{equation}\label{eq:gendec}
 u=\underbrace{\hat{u}(x,y,z,t)}_{\mathcal{O}(1)}+\underbrace{\tilde{u}(x,y,z,t)}_{\mathcal{O}(\varepsilon)},\,\,w=\underbrace{\hat{w}(x,y,z,t)}_{\mathcal{O}(1)}+\underbrace{\tilde{w}(x,y,z,t)}_{\mathcal{O}(\varepsilon)}.
\end{equation}
The leading-order contributions are required to satisfy \eqref{eq:genlwfl} truncated to $\mathcal{O}(1)$, but with additional $\mathcal{O}(\varepsilon)$ forcing terms, $K_u$ and $K_w$, which ensures that the local gap-averaged flow rates, $q_x$ and $q_z$, are determined entirely by $\hat{u}$ and $\hat{w}$. We thus have the following definitions:
\begin{equation}\label{eq:genWRIBLdec}
\begin{split}
\partial_y[\mu \, \partial_y\hat{u}]=-K_u,\qquad\qquad &\partial_y[\mu \, \partial_y\hat{w}]=-K_w,\\
\hat{u}(0)=\hat{u}(1)=0,\qquad\qquad &\hat{w}(0)=\hat{w}(1)=0,\\
\int_0^1{\hat{u}}\,dy=q_x,\qquad\qquad &\int_0^1{\hat{w}}\,dy=q_z.
\end{split}
\end{equation}
Here, $K_u$ and $K_w$ are $y$-independent functions, which are determined by solving the boundary value problems in \eqref{eq:genWRIBLdec} and applying the integral conditions. Thus, we obtain the leading order velocity profiles $\hat{u}$ and $\hat{w}$, which only depend on $x$, $z$ and $t$ through their dependence on $q_x$ and $q_z$, respectively. Note that $\hat{u}$ and $\hat{w}$ will be parabolic only if $\mu$ is uniform in the $y$ direction. Conversely, when $\mu$ is a function of $y$ then \eqref{eq:genWRIBLdec} will capture the consequent non-parabolic variation of velocity across the thin gap. 

The specification of the leading order velocity field is completed by using the continuity equation \eqref{eq:genlwcont} to determine the leading-order cross-gap velocity $\hat{v}$:
\begin{equation}\label{eq:genWRIBLv}
    \hat{v} = -\int_0^y{\left( \partial_{x} \hat{u}+\partial_{z} \hat{w} \right) dy},
\end{equation}
where we have used the no-penetration boundary condition, $\hat{v}|_{y=0}=0$.

Returning to the decomposition \eqref{eq:gendec}, we see that the definitions in \eqref{eq:genWRIBLdec} imply that the $\mathcal{O}(\varepsilon)$  corrections to the velocity field satisfy the following boundary and integral conditions:
\begin{equation}\label{eq:genlwdec}
\begin{split}
\tilde{u}(0)=\tilde{u}(1)=0,\qquad\qquad &\tilde{w}(0)=\tilde{w}(1)=0,\\
\int_0^1{\tilde{u}\,dy}=0,\qquad\qquad &\int_0^1{\tilde{w}\,}dy=0.
\end{split}
\end{equation}
The domain equations for the $\mathcal{O}(\varepsilon)$  corrections may be obtained by substituting \eqref{eq:gendec} and \eqref{eq:genWRIBLdec} into \eqref{eq:genlw}, but they are not needed to close the averaged model. 

We now proceed to average the governing equations, beginning with the continuity equation  \eqref{eq:genlwcont}, which on integrating across the thin gap and using the no-slip and no-penetration conditions yields the following \textit{exact} mass balance equation:
\begin{equation}\label{eq:genmassbal}
\partial_x q_x +\partial_z q_z=0 .
\end{equation} 

Next, we must average the momentum equations in order to obtain relations between the flow rate and the pressure gradient. In order to consistently close the $\mathcal{
O}(1)$ viscous term, however, we carry out a weighted average by multiplying \eqref{eq:genlwfl} with a weight function $F_v$ and then integrate across the gap, to yield (after substituting the velocity decomposition \eqref{eq:gendec}):
\begin{equation}\label{eq:genintfldec}
\begin{split}
&-\varepsilon\partial_z p\left[\int_0^1{F_v\,dy}\right] +\int_0^1\partial_y \left[\mu\,\partial_y(\hat{w}+\tilde{w})\right]F_v\,dy=0,\\&-\varepsilon\partial_x p\left[\int_0^1{F_v\,dy}\right] +\int_0^1\partial_y \left[\mu\,\partial_y(\hat{u}+\tilde{u})\right]F_v\,dy=0.
\end{split}
\end{equation}
The $O(\epsilon)$ corrections, $\tilde{u}$ and $\tilde{w}$, may be entirely eliminated from \eqref{eq:genintfldec} by using a Galerkin weight function defined as follows \citep{RQ2002}: 
\begin{equation}\label{eq:genweight}
\partial_y[\mu\,\partial_yF_v]=-1 \qquad \qquad \mbox{and}\qquad \qquad F_v(0)=F_v(1)=0.
\end{equation}
Now, on applying integration by parts twice to the right-most integrals of \eqref{eq:genintfldec} and using the boundary conditions \eqref{eq:genlwdec} and \eqref{eq:genweight}, we obtain the closed WRIBL flow model:
\begin{equation}\label{eq:genWRIBL}
\partial_x q_x +\partial_z q_z=0, \quad 	q_x=-\partial_x p\left[\int_0^1{F_v\,dy}\right], \quad q_z=-\partial_z p\left[\int_0^1{F_v\,dy}\right],
\end{equation}
where we have repeated the mass balance equation \eqref{eq:genmassbal} for completeness.

The cross-gap variation of viscosity, and the consequent non-parabolic local flow profile, enter the WRIBL model \eqref{eq:genWRIBL} through the weight function $F_v$, as well as through the leading-order velocity profiles $\hat{u}$ and $\hat{w}$. In fact, if $\mu$ is constant, then $\hat{u}$, $\hat{w}$, and $F_v$ become parabolic, and \eqref{eq:genWRIBL} reduces to the Darcy flow model. Note that while $\hat{u}$, $\hat{w}$ and $\hat{v}$ are not seen in \eqref{eq:genWRIBL}, they will appear in the convective terms of the conservation equation for the field variable that determines the viscosity (as it does in the energy balance equation in \S\ref{sec:TVFmathmodel}).

In this derivation, we have, for simplicity, retained terms only up to $\mathcal{O}(\varepsilon)$. However, the model can be further improved by retaining $O(\varepsilon^2)$ terms and using a similar weighted residual procedure to include the effects of mechanical inertia~\citep{RQ2012} and in-plane viscous diffusion, as demonstrated by \cite{Dietze2013,Dietze2015} for interfacial flows. (The leading-order velocity profiles $\hat{u}$ and $\hat{w}$ will then make an appearance in the averaged flow equations via these additional terms.) The resulting $\mathcal{O}(\varepsilon^2)$ WRIBL model would be an improved version of the inertial Darcy-Brinkmann equation, with corrections due to cross-gap variations in viscosity. We leave such extensions for future work, however, and turn to a specific application of the WRIBL flow model~\eqref{eq:genWRIBL}.

\section{WRIBL model for thermoviscous fingering}\label{sec:TVFmathmodel}

We now take up the problem of thermoviscous fingering which arises in the Hele-Shaw flow of a fluid with a viscosity that decreases sharply with increasing temperature. Figure~\ref{fig:schematic} presents a schematic of the Hele-Shaw flow, in which hot fluid, entering the channel at a mean temperature $T_H$, is driven primarily in the $z$ direction by a fixed pressure drop $P_{in}-P_{out}$. The temperature of the colder walls is assumed either to be constant, with a value $T_C$, or linearly decreasing along the flow direction, as $T_{H}-(T_H-T_C) z/L$. The former case has been studied before in the context of magma flows \citep{Wylie1995,Helfrich1995}, while the latter case, which is more relevant to industrial processes such as injection moulding, is considered here for the first time. In both cases, the fluid will lose heat to the colder walls and become cooler and more viscous as it flows. Importantly, the temperature and therefore the viscosity will vary across the gap and result in a non-parabolic cross-gap velocity profile \citep{Wylie1995}. It is this leading-order feature of the problem that is neglected by the Darcy model used in previous work \citep{Helfrich1995}, and which we seek to include using the WRIBL averaging scheme. 

This problem serves as an ideal test case because the dependence of viscosity on temperature is not merely incidental, but rather plays a key physical role in thermoviscous fingering. This channelling instability arises because advective heat supply, which competes with heat loss to the cold side-walls, depends on the viscosity and thus on the local temperature. A positive perturbation to the local temperature inside the fissure decreases the local viscosity, and therefore increases the velocity. This, in turn, decreases the residence time of the fluid parcel in the conduit, thus reducing heat loss to the cold walls and further increasing the local temperature. It is this positive feedback that underlies the fingering instability, which ultimately produces channels of low-viscosity, fast-moving magma separated by regions of high viscosity slow-moving magma. On the surface, at the outlet of the fissure, this instability manifests as a transition in the spatial form of the emerging flow: from a uniform sheet of magma to well-separated spouts.

To model this coupled flow and heat transport problem, we derive an averaged transport equation for temperature, which is then coupled to the averaged flow model. We use the WRIBL method for this purpose, and the derivation parallels that of the flow model (\S\ref{sec:mathmodel}), except that the situation is now simpler as we are dealing with a single scalar field and a constant diffusion coefficient. We begin with the energy equation in dimensional form:
\begin{equation}\label{eq:geqn}
\partial_{t^*}T^*+ \vec{v}^* \cdot\nabla^* T^*= \kappa\nabla^{*2}T^*,
\end{equation}
where, $\kappa$ is the thermal diffusivity. 

In case 1, we consider the temperature of both side walls $T_w^*$ to be constant,
\begin{equation}\label{eq:swbc1}
  T^*|_{y^*=0}=T^*|_{y^*=d}= T_w^* =T_C,
\end{equation} 
and assume that the fluid enters with a mean temperature $T_H$ and a parabolic cross-gap profile,
\begin{equation}\label{eq:swic1}
T^*|_{z^*=0} = T_C+ 6 (T_H-T_C) \frac{y^*}{d} (1-\frac{y^*}{d}).
\end{equation} 

\citet{Wylie1995} showed that if the fluid enters with a uniform temperature, then the temperature profile rapidly evolves, over a short entrance length, to a near-parabolic profile. This initial rapid evolution may be seen as a relaxation of the system from its initial condition to the low-dimensional slow-manifold on which the dynamics may be described by reduced-order models~\citep{Robertsbook,Roberts2017}. In general, this initial relaxation cannot be captured by averaged models, although error-minimizing strategies have been proposed~\citep{Balakotaiah2010,Roberts1992,Roberts2001}. Here, because our goal is to compare between the WRIBL and Darcy models, we avoid the singular initial condition of a uniform temperature and instead use the parabolic initial condition~\eqref{eq:swic1}.

In case 2, we enforce a linearly decreasing temperature profile along the walls,
\begin{equation}\label{eq:swbc2}
    T^*|_{y^*=0}=T^*|_{y^*=d}=T_w^* =T_{H}-(T_H-T_C) \frac{z}{L},
\end{equation} 
and consider the temperature of the fluid at the inlet to be the same as the local wall-temperature, which implies that
\begin{equation}\label{eq:swic2}
T^*|_{z^*=0} = T_H.
\end{equation} 
In contrast to case 1, the fluid in case 2 starts out in thermal equilibrium with the walls, and so the issue of a rapid initial relaxation does not arise. This contradistinction is why we find it instructive to consider both cases.

To scale these equations, we use \eqref{eq:scaling} along with the nondimensional temperature,
\begin{equation*}
T=\frac{T^*-T_C}{T_H-T_C},
\end{equation*}
and on dropping terms of order higher than $\varepsilon$ we obtain the following scaled heat transport equation:
\begin{equation}\label{eq:lwtheta}
\varepsilon \left(\partial_tT+ u\partial_xT+ v\partial_yT+ w\partial_zT \right)= \partial_y^2T.
\end{equation}
Here, we have chosen the characteristic velocity to be $U=\kappa/d$, which amounts to setting the P\'eclet number $Pe=Ud/\kappa$ to unity, in contrast to $\Rey$ which was taken to be $\mathcal{O}(\varepsilon)$ or smaller in \S\ref{sec:mathmodel}. Thus, we retain thermal inertia while neglecting momentum inertia, a choice that is motivated by the fact that in magma flow through basaltic fissures, $Pe$ is typically much larger than $\Rey$  \citep{Wylie1995,Wylie1999}. Moreover, as discussed above, convective heat transport plays a key role in the positive feedback between flow rate and temperature perturbations which underlies the fingering instability. 

Equation \eqref{eq:lwtheta} is accompanied by the following non-dimensional inlet and boundary conditions, which differ for the two cases.\\
Case 1:
\begin{subequations}\label{eq:lwthetabc}
\begin{equation}\label{eq:swicnd1}
  T|_{z=0} = 6 y (1-y),
\end{equation} 
\begin{equation}\label{eq:swbcnd1}
  T|_{y=0}=T|_{y=1}=T_w=0.
\end{equation} 
Case 2:
\begin{equation}\label{eq:swicnd2}
  T|_{z=0} = 1,
\end{equation} 
\begin{equation}\label{eq:swbcnd2}
  T|_{y=0}=T|_{y=1}=T_w=1-z.
\end{equation} 
\end{subequations}

We now employ the WRIBL technique to obtain an averaged heat transfer equation. The derivation closely follows that in \S\ref{sec:mathmodel}, and so we only present the key steps here. First, the temperature field is split into the wall temperature $T_w$ and an internal variation which in turn is decomposed into $\mathcal{O}(1)$ and $\mathcal{O}(\varepsilon)$ components:
\begin{equation}\label{eq:dec}
T=T_w+\underbrace{\hat{T}(x,y,z,t)}_{\mathcal{O}(1)}+\underbrace{\tilde{T}(x,y,z,t)}_{\mathcal{O}(\varepsilon)}\,\,
\end{equation}
The leading order component of the internal variation is defined as:
\begin{equation}\label{eq:WRIBLdec}
\partial^2_y\hat{T}=-K_{\theta},\quad \hat{T}(0)=\hat{T}(1)=0, \quad \int_0^1{\hat{T}}dy=\theta-T_w,
\end{equation}
where $K_{\theta}$ is completely determined in terms of the thin-gap averaged temperature $\theta (z,x,t)$ and the wall temperature $T_w(z)$. Thus, the leading order cross-gap temperature profile $\hat{T}(y;\theta)$ arises from a competition between thermal conduction and convection and takes on a parabolic form, parameterized by $\theta$ and $T_w$. The $\mathcal{O}(\varepsilon)$ correction in \eqref{eq:dec} must satisfy the following conditions:
\begin{equation}\label{eq:lwdec}
 \tilde{T}(0)=\tilde{T}(1)=0, \quad \int_0^1{\tilde{T}}dy=0
\end{equation}


Next, we substitute the temperature decomposition \eqref{eq:dec} along with the velocity decomposition \eqref{eq:gendec} from \S\ref{sec:mathmodel} into \eqref{eq:lwtheta} and neglect all terms of $O(\varepsilon\tilde{u}) \sim O(\varepsilon^2)$ and smaller to  obtain
\begin{equation}\label{eq:lwthetadec}
\varepsilon \left(\partial_t\hat{T}+ \hat{u}\partial_x\hat{T}+ \hat{v}\partial_y\hat{T}+ \hat{w}\partial_z\hat{T} + \hat{w}\partial_z T_w \right)= \partial_y^2(\hat{T}+\tilde{T})
\end{equation}
In order to derive a closed averaged model, we take the Galerkin weighted residual of \eqref{eq:lwthetadec} using a weight function $F_{\theta}$, defined as
\begin{equation}\label{eq:weight}
\partial^2_yF_{\theta}=-1,\qquad F_{\theta}(0)=F_{\theta}(1)=0.
\end{equation}
On using integration by parts, along with the boundary and integral conditions in \eqref{eq:WRIBLdec}, \eqref{eq:lwdec} and \eqref{eq:weight}, we exactly close the diffusive term on the right-hand-side of \eqref{eq:lwthetadec} and thereby obtain
\begin{equation}\label{eq:WRIBLtheta}
\varepsilon \left[\int_0^1F_\theta\left(\partial_t\hat{T}+\hat{u}\partial_x\hat{T}+\hat{v}\partial_y\hat{T}+\hat{w}\partial_z\hat{T} +\hat{w}\partial_z T_w \right)dy\right]=-(\theta-T_w)
\end{equation}

It remains to specify the functional dependence of the viscosity on temperature. A simple relationship that has been used in previous work \citep{Helfrich1995} is $\mu^*(T^*)=\mu_0\exp[\gamma^*(T_H-T^*)]$, where the characteristic viscosity $\mu_0 = \mu^*(T_H)$. In non-dimensional terms, we have
\begin{equation}
\mu=\exp[\gamma(1-T)]. 
\label{eq:muexpfull}
\end{equation}
Now, replacing $T$ with its leading order approximation, we obtain the following relationship which couples the averaged flow ~\eqref{eq:genWRIBL} and heat transport~\eqref{eq:WRIBLtheta} equations:  
\begin{equation}\label{eq:muexp}
\mu=\exp\left[\gamma\left(1-(T_w+\hat{T})\right)\right]. 
\end{equation}
Using \eqref{eq:muexp}, we solve \eqref{eq:genWRIBLdec} for $\hat{u}$ and $\hat{w}$, \eqref{eq:genWRIBLv} for $\hat{v}$, as well as \eqref{eq:genweight} for the weight function $F_v$. We then evaluate the integrals appearing in the averaged equations of flow \eqref{eq:genWRIBL} and heat transport \eqref{eq:WRIBLtheta} to obtain the coupled WRIBL model in terms of the gap-average temperature and the fluid flow rates:
\begin{subequations}\label{eq:WRIBLcomb}
\begin{equation}\label{eq:WRIBLcombF}
\partial_x q_x +\partial_z q_z=0,\qquad q_x=-\frac{F_1}{\mu(\theta)}\partial_x p,\qquad q_z=-\frac{F_2}{\mu(\theta)}\partial_z p,
\end{equation}
\begin{equation}\label{eq:WRIBLcombT}
\partial_{t}\theta+ G_1\; q_x \partial_{x}\theta +G_2\;q_z \partial_{z}\theta=-G_3 \;(\theta-T_w).
\end{equation}
The expressions for $F_i$ and $G_i$, which are functions of $q_z$, $q_x$, $\theta$, and $T_w$, are given in a Mathematica\textsuperscript{\textregistered} notebook in the Other supplementary material. Note that for a constant viscosity $F_i = 1/12$ and \eqref{eq:WRIBLcombF} reduces to the Darcy flow equation. 

The inlet condition for the cross-gap averaged temperature is 
\begin{equation}\label{eq:WRIBLinlet}
\theta|_{z=0} = 1.
\end{equation}
\end{subequations}
Equations \eqref{eq:WRIBLcomb} constitute the coupled WRIBL model. It is distinguished from previously used Darcy models, such as the one described in the next section, by the nonlinear functions $F_i$ and $G_i$ through which the cross-gap variations in temperature and viscosity enter the problem. 

\section{Darcy model for thermoviscous fingering}
\label{sec:Darcy}

We now briefly describe the Darcy-like model used by \cite{Helfrich1995}, which unlike our WRIBL model neglects cross-gap variations in viscosity. The cross-gap averaged flow rates are described by Darcy's law,
\begin{subequations}\label{eq:Darcy}
\begin{equation}\label{eq:Darcyfl}
\partial_x q_x +\partial_z q_z=0,\qquad 	q_x=-\frac{1}{12\mu(\theta)}\partial_x p,\qquad q_z=-\frac{1}{12\mu(\theta)}\partial_z p,
\end{equation}
where the viscosity $\mu (\theta) = \exp[\gamma(1-\theta)]$ varies with the cross-gap averaged temperature $\theta (z,x,t)$, which is in turn governed by
\begin{equation}\label{eq:Darcytheta}
\varepsilon\left(\partial_{t}\theta+ q_x \partial_{x}\theta +q_z \partial_{z}\theta\right)=-\pi^2(\theta-T_w),
\end{equation}
along with the inlet condition,
\begin{equation}\label{eq:Darcyinlet}
\theta|_{z=0} = 1.
\end{equation}
Recall that $T_w = 0$ for case 1 whereas $T_w = 1-z$ for case 2.
\end{subequations}

 To better appreciate the differences between this Darcy model and our WRIBL model, we shall attempt to reduce the WRIBL model \eqref{eq:WRIBLcomb} to the Darcy model \eqref{eq:Darcy}. Firstly, one must neglect the $O(1)$ variation of temperature and therefore of viscosity which leads to $F_i = 1/12$ and reduces \eqref{eq:WRIBLcombF} to \eqref{eq:Darcyfl}. The disregard of cross-gap temperature variations amounts to assuming that $\hat{T}=\theta$ and choosing the Galerkin weight function $F_\theta = 1$, both of which result in $G_1=G_2 =1$. The convective terms of \eqref{eq:WRIBLcombT} then simplify to the convective terms on the left-hand-side of \eqref{eq:Darcytheta}. However, we also obtain $G_3 =0$ which implies an absence of heat transfer to the side walls; the loss of this key physical ingredient is a direct outcome of assuming the temperature to be uniform across the thin gap. In \citet{Helfrich1995}, the Darcy model is salvaged by introducing a heuristic heat loss term, $-\pi^2(\theta-T_w)$, which then yields \eqref{eq:Darcytheta}. This sink term is the conductive heat loss that would occur if the cross-gap temperature profile is that of the leading eigenfunction of the heat diffusion equation, that is, $(\theta-T_w) (\pi/2) {\rm sin}(\pi y)$ [note that $T_w = 0$ in \citet{Helfrich1995}]
 
 In summary, to obtain the Darcy-like model~\eqref{eq:Darcy}, two asymptotically unjustified simplifications must be made: (a) neglect the leading order, cross-gap variation of temperature, (b) artificially introduce a heat sink term to model the conductive heat loss to the wall. The WRIBL strategy, presented in \S\ref{sec:mathmodel} and \S\ref{sec:TVFmathmodel}, allows us to retain all leading order effects and yet derive an averaged model whose computational simplicity is comparable to that of the Darcy-like model.

\section{Testing the WRIBL model: importance of cross-gap variations}\label{sec:results}

We evaluate the relative accuracy of the WRIBL model, and the importance of cross-gap temperature and viscosity variations, by comparing the predictions of the WRIBL model with that of the Darcy model of \cite{Helfrich1995}. In addition, we solve the the full 3D model, given by \eqref{eq:genlw} and \eqref{eq:lwtheta}-\eqref{eq:lwthetabc}, and consider its results as the `truth' against which both average models are tested. This 3D model was first studied by \cite{Wylie1995} for the case of constant wall temperature $T_w = 0$. 

We focus on predictions of the uniform base flow state, its multiplicity, and its linear stability. The ordinary differential equations of the two averaged models are integrated using the adaptive-stepping \textit{NDSolve} subroutine in Mathematica\textsuperscript{\textregistered} v.10.3.. The partial differential equations of the full 3D model are solved numerically by discretizing the thin gap direction using the Chebyshev pseudo-spectral method and using the implicit integration scheme described in \cite{Wylie1995} for evolving the temperature along the flow direction.

We now proceed to compare the averaged models, first considering a constant cold temperature at the walls (case 1) and next adopting a linearly decreasing wall temperature (case 2). In both cases, we begin with the prediction of the steady and uniform base flow and then move on to its linear stability characteristics.

\subsection{Constant wall temperature}

\subsubsection{Uniform base state flow}

To calculate the base flow, we discard derivatives in time (steady state) and in the lateral $x$-direction (uniform flow) and set the lateral velocity component to zero: $\bar{u} = 0$, where the overbar indicates a base state variable. The latter restriction implies that $\bar{q}_x=0$ and thus, from \eqref{eq:genmassbal}, that $\bar{q}_z$ is constant. Now, while we assume that the flow occurs under constant pressure-drop conditions, the computation of the base flow states is eased by beginning with a specified value of $\bar{q}_z$ and then solving for the flow field to obtain the pressure drop $\overline{\Delta P}$. However, when interpreting the results, it is important to remember that $\overline{\Delta P}$ is the input and $\bar{q}_z$ is the output.


 \begin{figure}
\centering
	\subfloat[]{\includegraphics[width=0.49\textwidth]{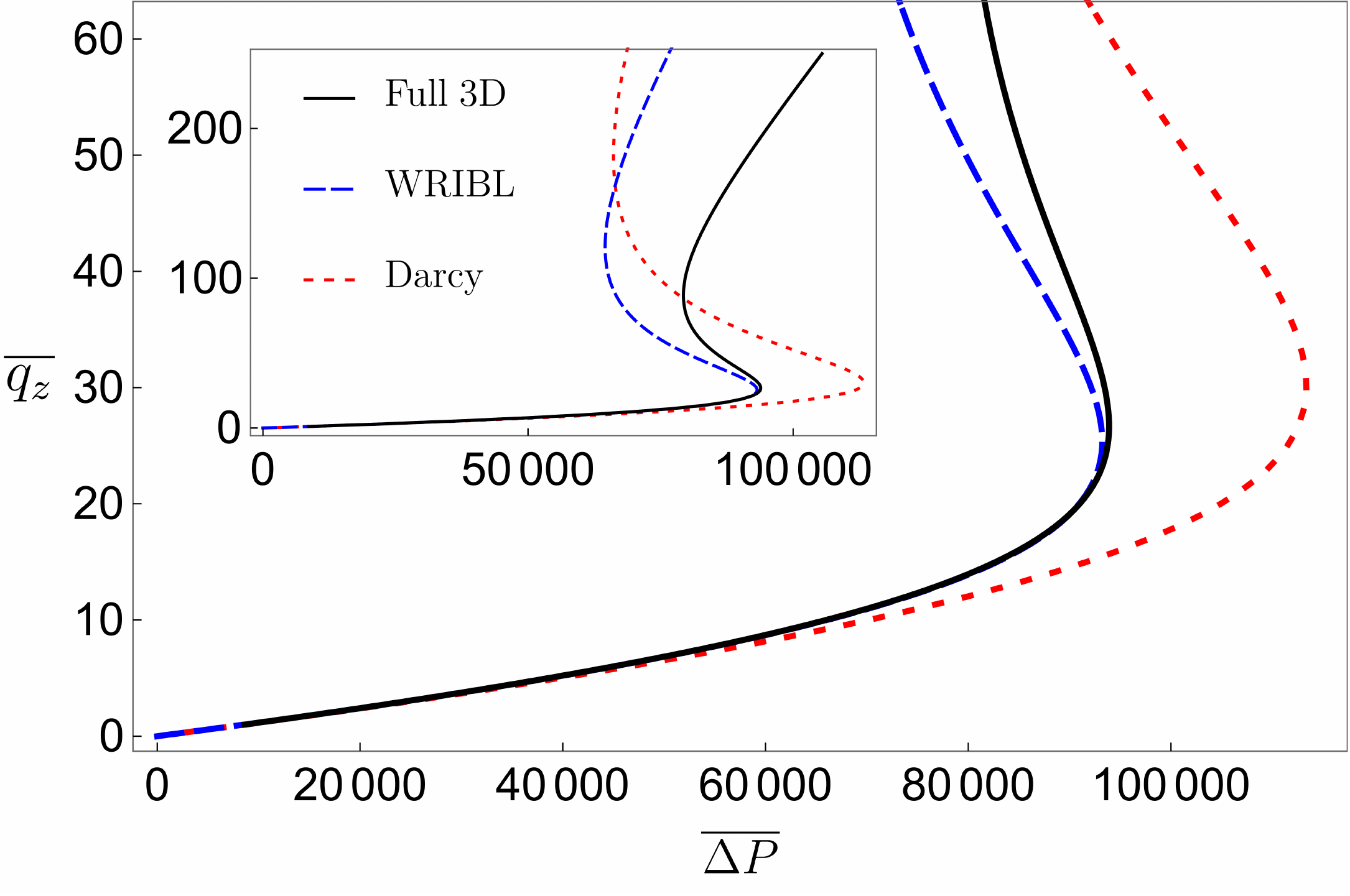}\label{fig:base}}\hspace{1pt}
    \subfloat[]{\includegraphics[width=0.49\textwidth]{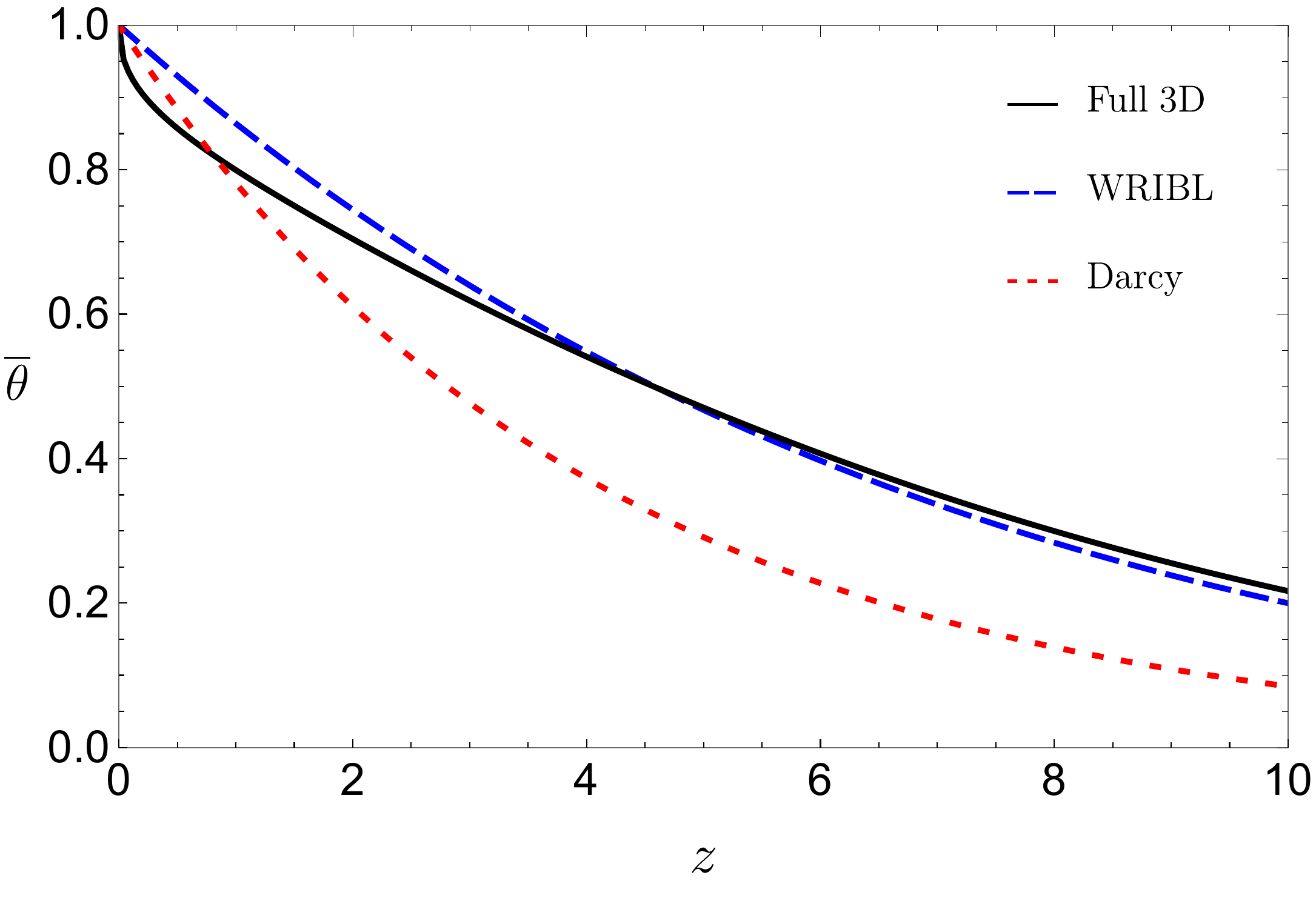}\label{fig:basetheta}}
     \caption{(a) The base state flow rate ($\overline{q}_z$) as a function of the applied pressure drop ($\overline{\Delta P}$), as predicted by all three models (see the legend), for the case of a constant wall temperature. The inset shows the behaviour over a wide range of $\overline{q}_z$, while the main panel focuses on lower values of $\overline{q}_z$. (b) Evolution of the gap-averaged base state temperature ($\overline{\theta}$) along the primary flow direction ($z$), for the state with $\overline{q}_z$=40. Parameter values: $L=10$ and $\gamma=4.3$}
\end{figure}

Figure \ref{fig:base} presents a typical flow rate -- pressure drop curve, as predicted by the 3D equations (black-solid), and the WRIBL (dashed-blue) and Darcy (dotted-red) averaged models. From the inset we see that all three models exhibit S-shaped curves, although a closer resemblance to the overall shape of the 3D-model curve is found in the WRIBL curve. All three curves predict the existence of multiple steady states over a range of pressure-drop values (bounded by the two turning points). The lower branch corresponds to a slow moving cold flow, while the upper branch corresponds to a fast moving hot flow. The backward branch with $d\bar{q_z}/d\overline{\Delta P}<0$ is unstable \citep{Helfrich1995,Wylie1995}, as discussed further below.

Quantitatively, the predictions of the three models coincide only for low flow rates, as seen more clearly in the main panel of figure \ref{fig:base}. For states with higher flow rates, convective effects are prominent and the averaged models exhibit errors which increase significantly beyond the first turning point. Note that within our non-dimensional framework, the flow rate $q_z$ may be interpreted as an effective P\'eclet number, representing the relative importance of (longitudinal) convective and (cross-gap) conductive thermal transport. So, averaging methods, based on relatively fast diffusion across the thin gap, are bound to fail at high flow rates. We therefore restrict our further comparison of the two averaged models to relatively low flow rates. In this regime, the WRIBL model is far more accurate than the Darcy model (cf. figure \ref{fig:base}); its prediction of the flow rate matches exactly with that of the full 3D model up to flow rates of about $\bar{q}_z=20$, whereas the Darcy model works well only till $\bar{q}_z=10$. The deviation of the WRIBL curve from the 3D result at higher flow rates is also rather moderate in comparison with the Darcy result.

\begin{figure}
\centering
	\subfloat[]{\includegraphics[width=0.48\textwidth]{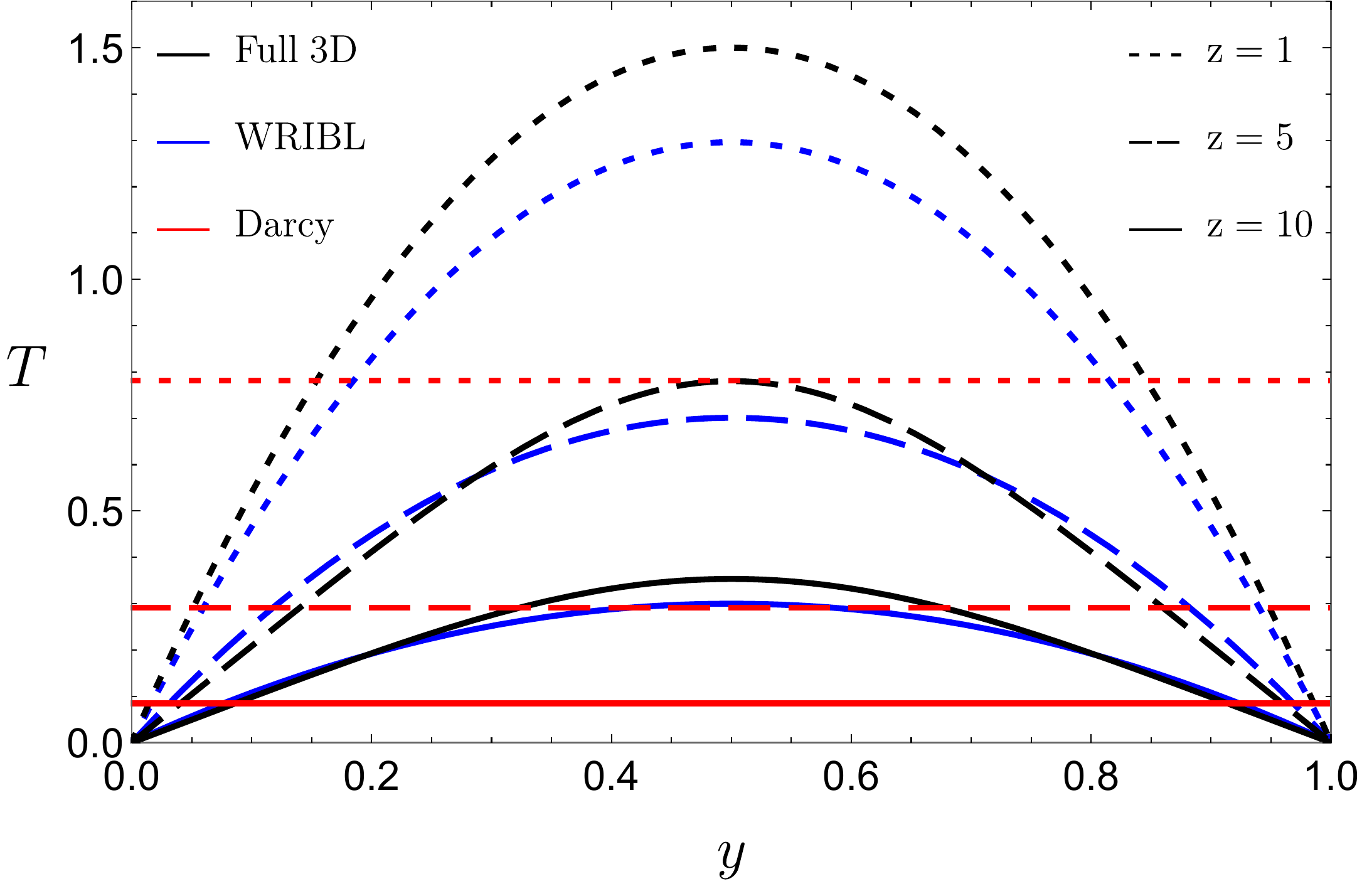}\label{fig:temp_comp}}\hspace{1pt}
    \subfloat[]{\includegraphics[width=0.48\textwidth]{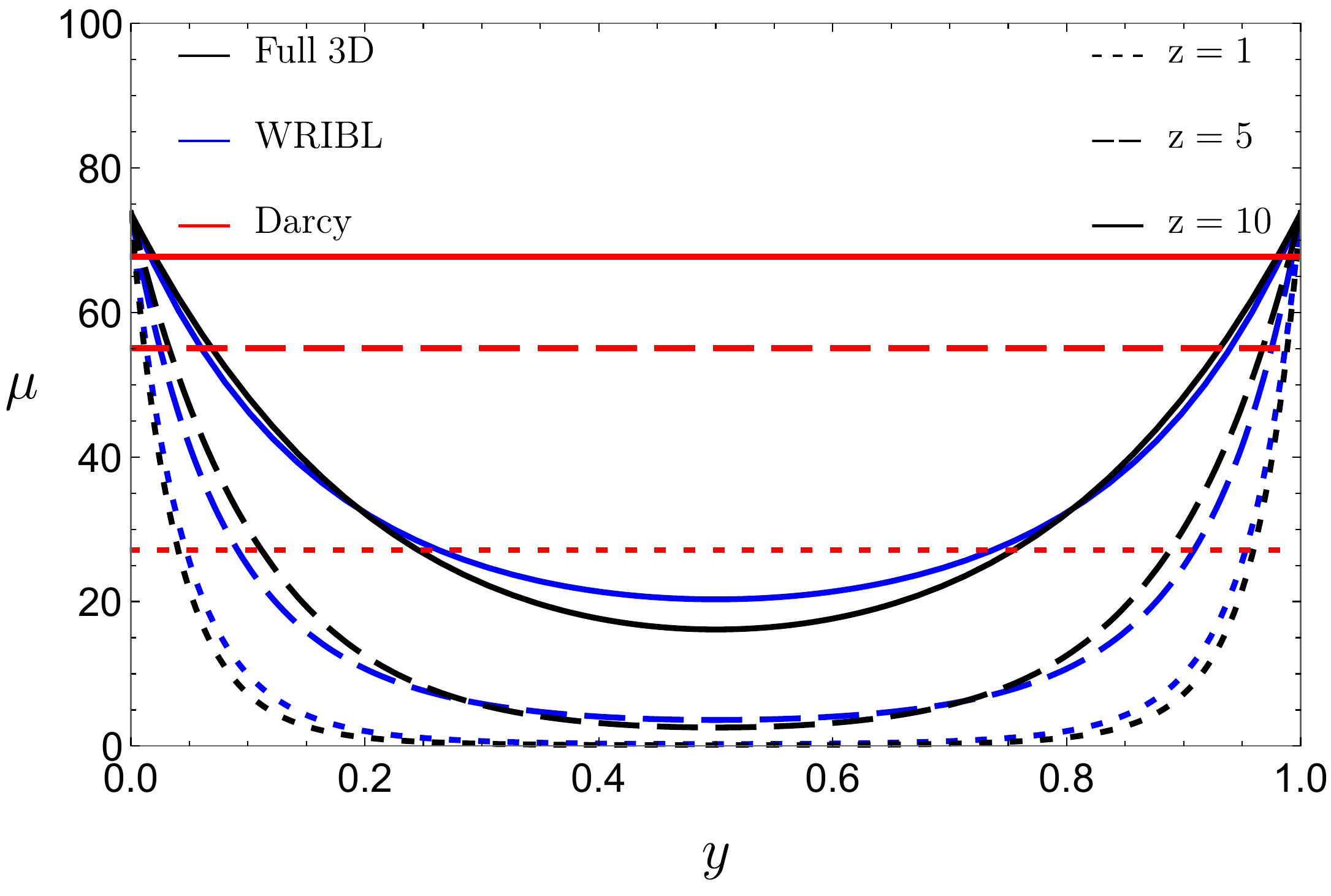}\label{fig:mu_comp}}\hspace{1pt}
    \subfloat[]{\includegraphics[width=0.48\textwidth]{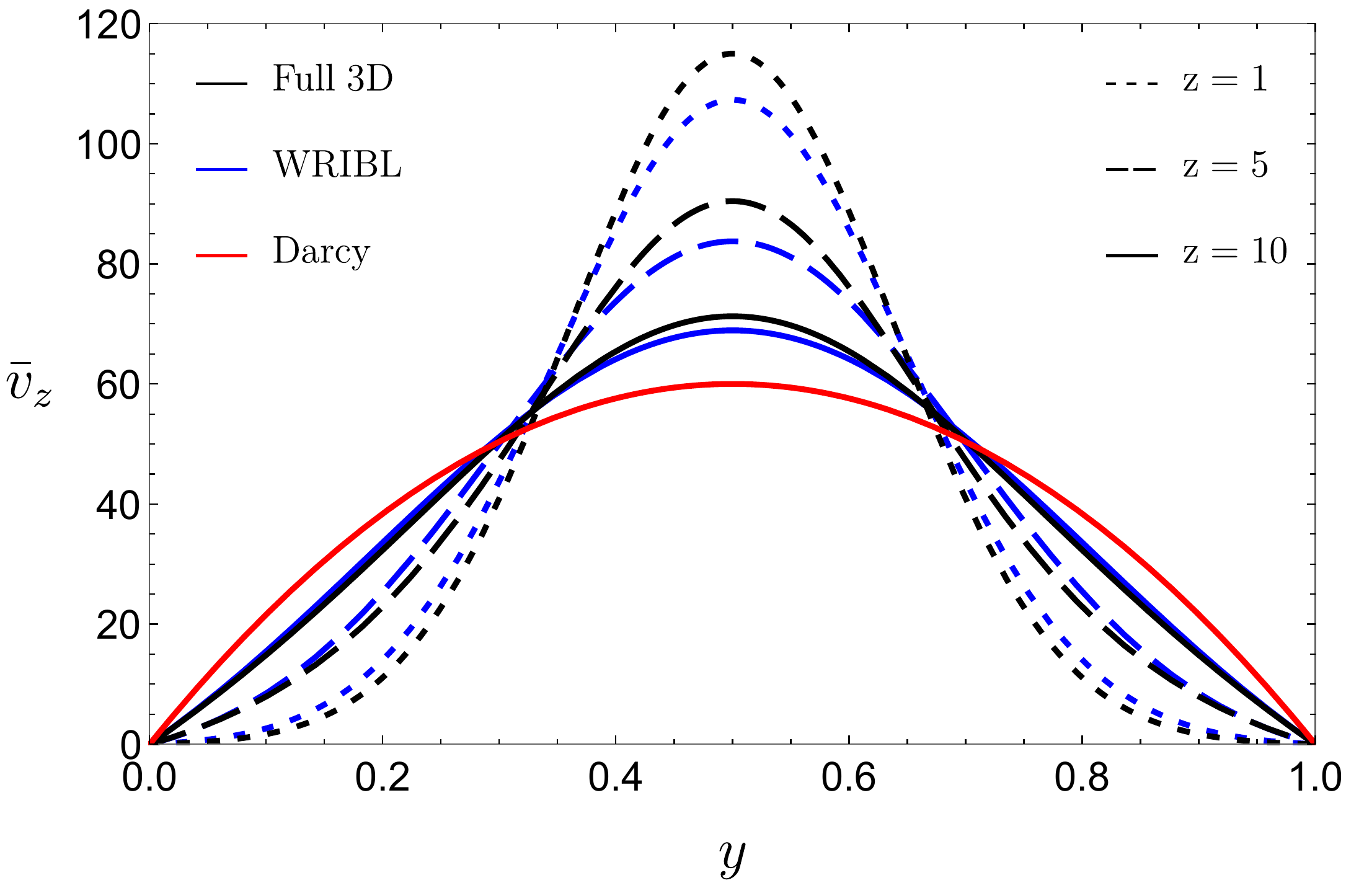}\label{fig:w_comp}}
     \caption{Base state profiles, across the thin gap, of (a) temperature, (b) viscosity, and (c) the longitudinal component of velocity, at different locations along the channel: $z$=1 (dotted), $z$=5 (dashed) and $z$=10 (solid). The predictions of the 3D, WRIBL and Darcy models are shown in black, blue and red, respectively (see the legend). Parameter values: $\gamma=4.3$, $\overline{q}_z$=40.}
     \label{fig:crossgap}
\end{figure}

Let us now focus on the state corresponding to $\bar{q}_z = 40$ and examine the evolution of the temperature of the fluid as it flows along the conduit. Figure \ref{fig:basetheta} compares the spatial evolution of the cross-gap average temperature obtained from the three models. Here too, we see that the WRIBL model performs much better than the Darcy model. Indeed, the WRIBL model captures the variation of the average temperature well, except for the inlet region where the relatively large difference between the mean temperature of the hot entering fluid and the temperature of the cold walls causes rapid heat loss to the walls.

The reason the Darcy model fares poorly is that it does not capture cross-gap variations in temperature and the resulting modulations of the viscosity and velocity. This fact is illustrated in Figure~\ref{fig:crossgap}, wherein panel (\textit{a}) compares the cross-gap temperature profiles predicted by the three models, at various longitudinal locations (see the legend). The Darcy model, of course, has a flat-line profile (red) corresponding to the local average temperature. In contrast, the parabolic, leading-order $\hat{T}$ profile of the WRIBL model (blue) matches relatively well with the profile of the full 3D model (black).

Figure~\ref{fig:mu_comp} contrasts the viscosity variation that results from the non-uniform cross-gap temperature, in the 3D and WRIBL models, with the non-varying viscosity assumed by the Darcy model. As expected, colder fluid near the walls has a much higher viscosity than the fluid at the centre of the thin gap---a fact that is entirely neglected by the Darcy model. Figure~\ref{fig:w_comp} reveals the impact of this varying viscosity on the velocity profile. In the constant-viscosity Darcy model the profile is always parabolic and therefore non-varying in the flow direction ($q_z$ is constant). In contrast, in the full 3D model the velocity profile changes quite dramatically: starting from a bell-shape near the inlet, it flattens out downstream as the cross-gap viscosity variation reduces (cf. figure~\ref{fig:mu_comp}). However, even at the outlet the profile is sharp-nosed compared to a parabola (red curve). Importantly, where the Darcy model fails, the WRIBL model works well and captures the key features of these cross-gap variations.

\begin{figure}
  \centerline{\includegraphics[width=9cm]{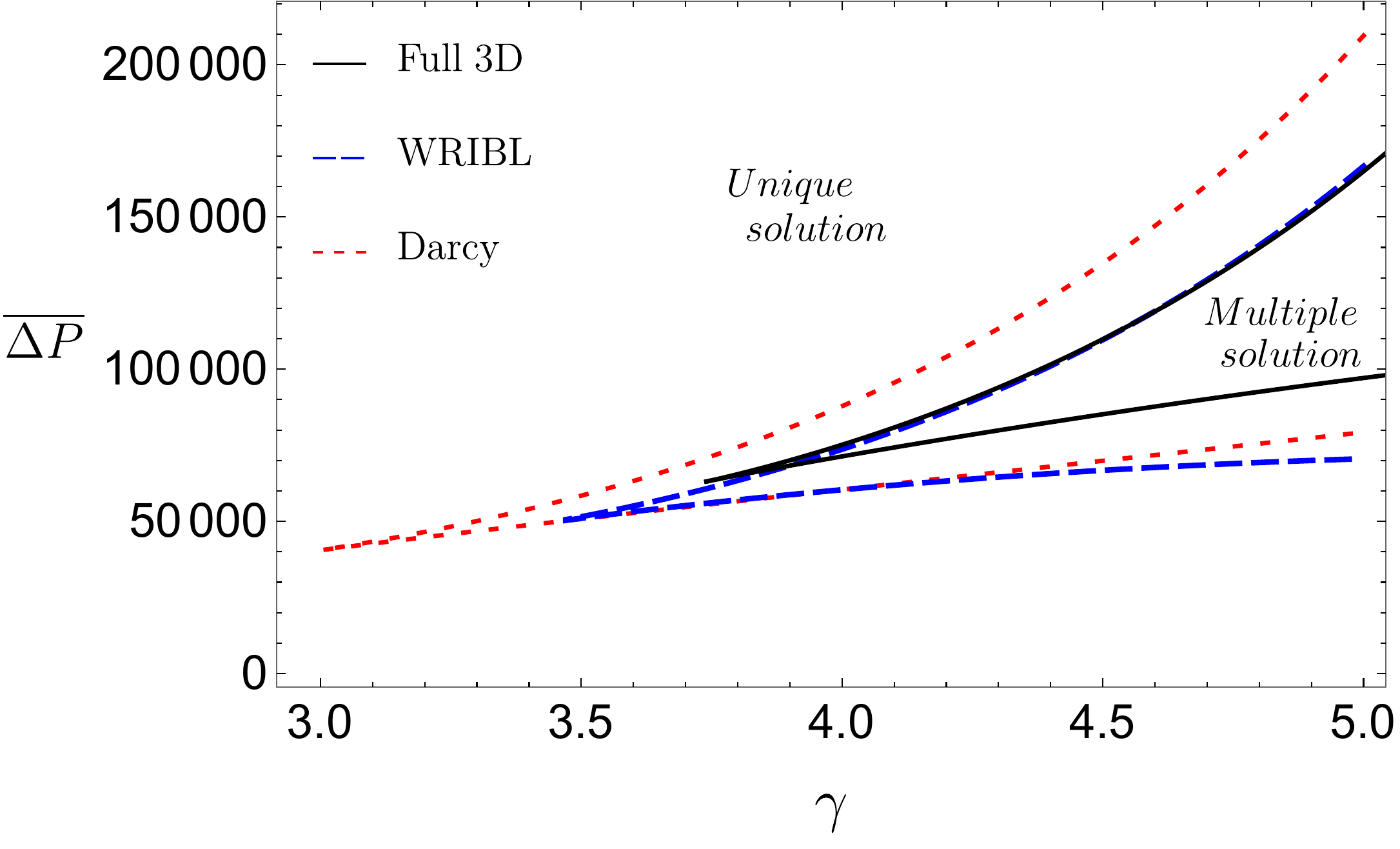}}
  \caption{Boundary between the zones of multiple and unique steady states in the $\gamma-\overline{\Delta P}$ parameter plane, as obtained from the predictions of the three models (see the legend). Parameter values: $L=10$.}
\label{fig:space}
\end{figure}

Returning to the issue of multiplicity, it is interesting to determine the region of the $\gamma-\overline{\Delta P}$ parameter space where multiple steady states are encountered in the different models. This region is depicted in figure~\ref{fig:space}, which shows the boundary between the zone of unique states, occurring at small $\gamma$, from the zone of multiple states, found at high $\gamma$. Increasing $\gamma$ increases the sensitivity of viscosity to temperature changes, and thus strengthens the positive nonlinear feedback between increments of temperature and flow rate. The cusp-like shape of the boundary, predicted by all three models, shows that unique states exist under both low and high pressure-drop conditions, which correspond to low and high flow rates, respectively. In the former case, thermal diffusion is dominant, while in the latter case convection is dominant. So, as discussed in \citet{Helfrich1995}, multiplicity (and the linear instability of the backward branch) only arises when convective heat supply competes with diffusive heat loss to the walls.

On comparing the boundaries predicted by the three models, we again find that the 3D results (black-solid) are much more closely approximated by the WRIBL model (blue-dashed) than the Darcy model (red-dotted). The fact that the Darcy boundary is shifted to smaller values of $\gamma$ compared to the 3D result was anticipated by \citet{Helfrich1995}, who pointed out that the cooling of the fluid near the walls results in a reduced effective sensitivity of the gap-averaged viscosity to the gap-averaged temperature.

\subsubsection{Linear instability}

Having examined the predictions of the uniform base flow, we now turn to its linear stability characteristics. For this, we perturb the base state using infinitesimal normal modes with lateral wavenumber $k$ and growth rate $\sigma$. So, for example, the average temperature in the WRIBL and Darcy models is expanded as follows:
\begin{align}\label{eq:normmode}
    \theta(x,z,t)&=\bar{\theta}(z)+\delta\; \theta'(z)\exp[\sigma t+ i k x],
\end{align}
where $\delta<<1$ and the prime denotes a perturbation. The flow rate and pressure are also perturbed in a similar manner, as are all the field variables in the full 3D equations. For each model, we then linearize the equations and obtain an eigenvalue problem for $\sigma$ with $k$ as a parameter. We follow the calculation procedures outlined in \citet{Helfrich1995}, for the averaged models, and in \citet{Wylie1995}, for the 3D model. 

The stability characteristics of the Darcy and 3D models are known to be in qualitative agreement, as shown by \citet{Helfrich1995} and \citet{Wylie1995}. In both cases, the backward branch of base states (cf. figure~\ref{fig:base}) is certainly unstable, with a growth rate that is positive over a range of wavenumbers starting with $k=0$ (uniform disturbances). However, the onset of instability does not coincide with the turning point, but rather occurs before it, on the lower branch of base states. The growth rate for these lower-branch unstable states is positive over a range of finite wavenumbers, bounded away from zero. These are the states that will likely give rise to sustained thermoviscous channelling, because the fastest growing normal mode has a non-zero wavenumber. In contrast, most of the backward branch states exhibit a growth-rate curve that peaks at $k=0$ and so rather than forming channels these states could simply transition to one of the two uniform states. The upper branch is found to be linearly stable to disturbances of all wavenumbers.

\begin{figure}
\centering
	\subfloat[]{\includegraphics[width=0.48\textwidth]{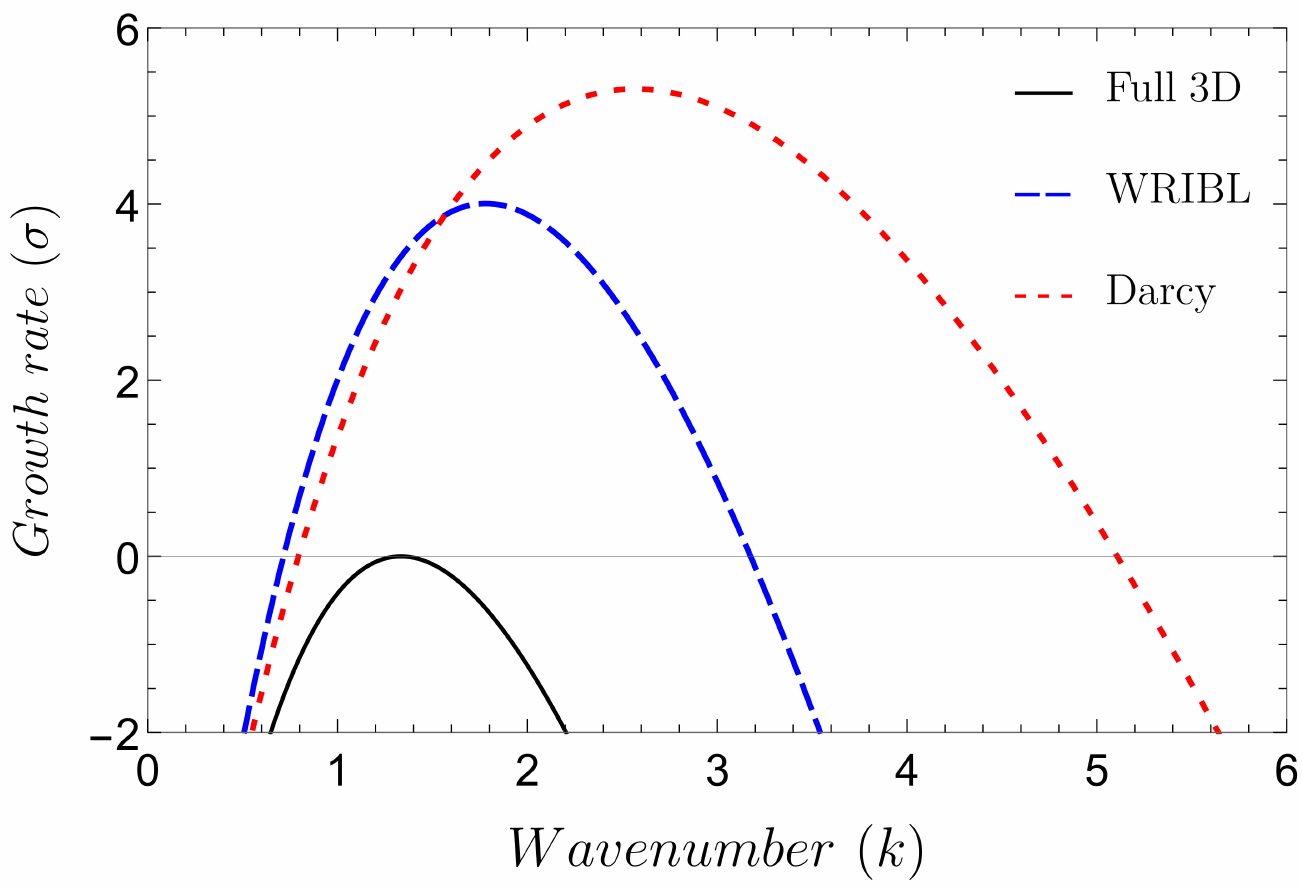}\label{fig:lsaq10}}\hspace{1pt}
     \subfloat[]{\includegraphics[width=0.48\textwidth]{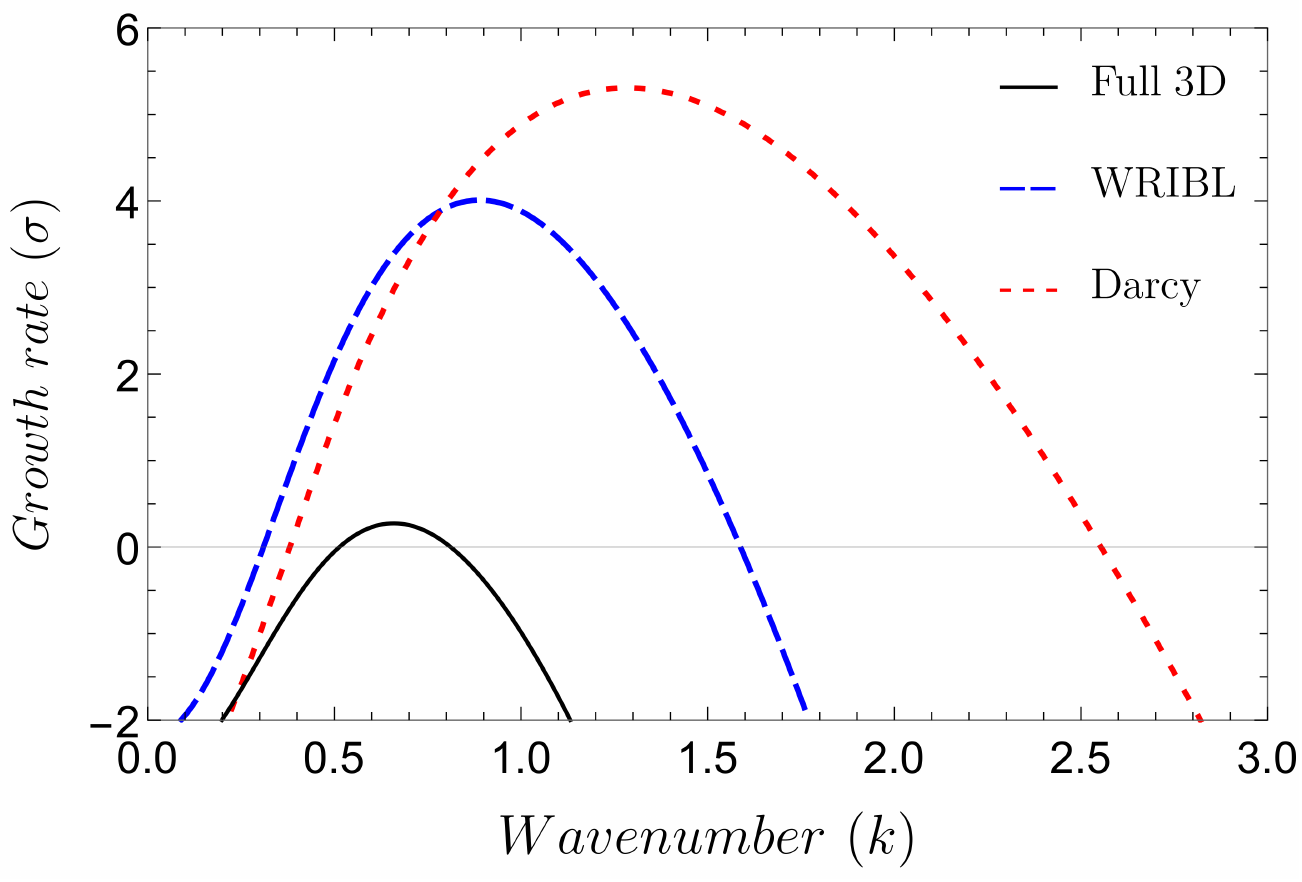}\label{fig:lsaq20}}
     \hspace{1pt}
     \subfloat[]{\includegraphics[width=0.48\textwidth]{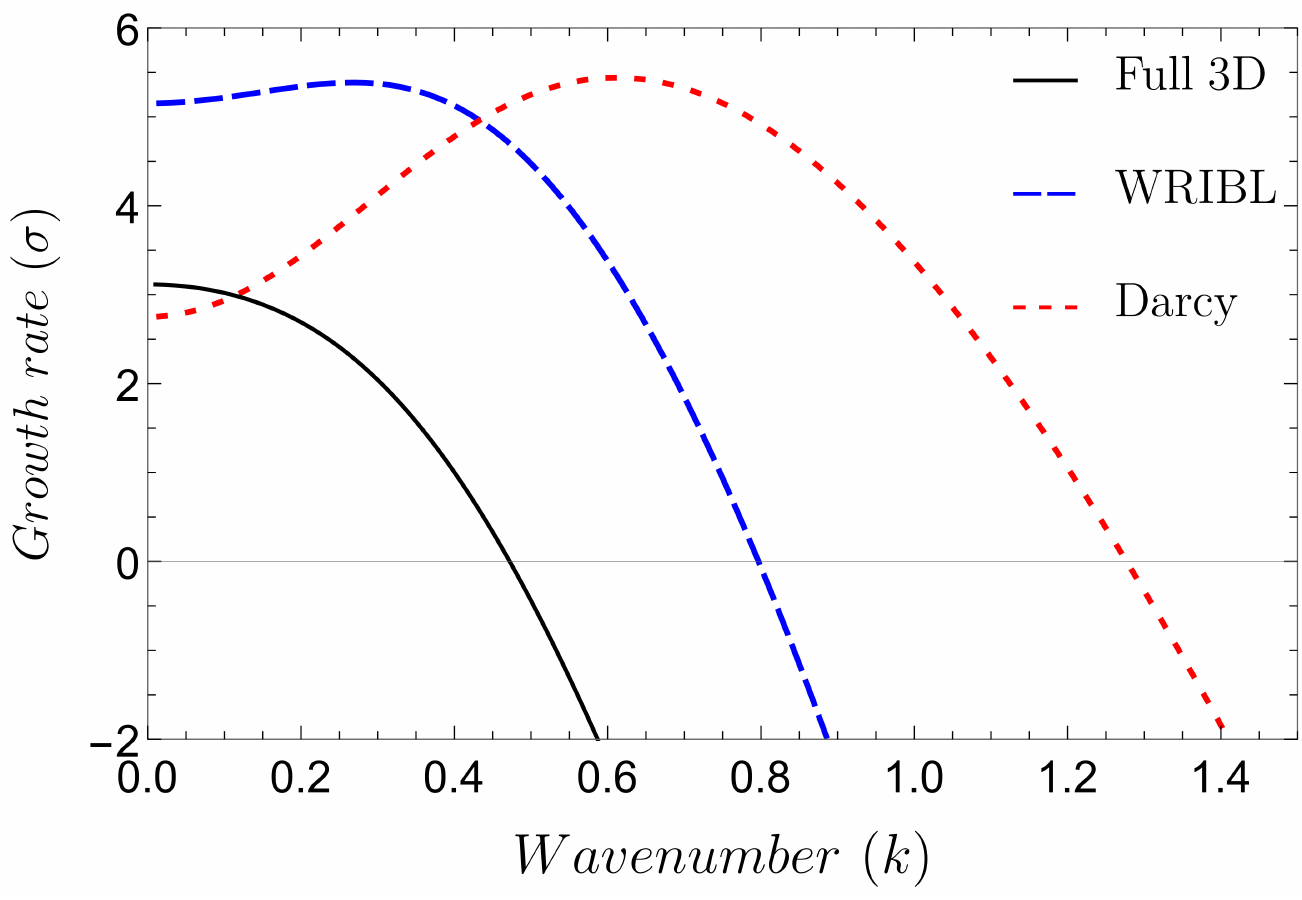}\label{fig:lsaq40}}
     \caption{Comparison of the linear stability results of the three models (see the legend) for the case of a constant wall temperature. The growth rate is plotted as a function of the wavenumber of the perturbation, for three different base states corresponding to flow rates of (a) $\bar{q}_z=10$, (b) $\bar{q}_z=20$, (c) $\bar{q}_z=40$. Parameter values: $L=10$, $\gamma=4.3$. }\label{fig:lsa1}
\end{figure}

Let us now compare the growth rates predicted by the three models quantitatively. This is done in Figure~\ref{fig:lsa1} for three different base states, labelled by their unique flow rate values. Figure~\ref{fig:lsaq10} corresponds to $\bar{q}_z=10$ which lies on the lower branch for all models. While the 3D model is at its onset of instability, both averaged models are already beyond their respective onsets and indeed exhibit a range of unstable wavenumbers. This over-prediction of the growth rates by the averaged models is also observed for the higher flow rate states of $\bar{q}_z=20$ and $\bar{q}_z=40$, shown in figure~\ref{fig:lsaq20} and \ref{fig:lsaq40} respectively. The last case lies on the backward branch for all three models, and so all three growth-rate curves in figure~\ref{fig:lsaq40} have positive growth rates for $k=0$. Again, the 3D results are better approximated by the WRIBL model, in terms of both the range of unstable wavenumbers and the shape of the growth-rate curve.

At this stage, the next natural step would be to carry out nonlinear simulations of the models to ascertain whether stable thermoviscous channels develop as a consequence of the linear instability. However, nonlinear simulations of the full 3D equations are beyond the scope of this work, and because our goal is to evaluate the accuracy of the averaged models relative to the 3D model we leave this issue for future work. It should be noted however that the question of whether stable channels form in this system remains unsettled. \citet{Wylie1995} used a continuation method on the full 3D model to search, just beyond the onset of instability, for steady three-dimensional states. They found none. In contrast, \citet{Helfrich1995} carried out transient simulations using the Darcy model and showed that stable hot-cold channels emerge for $\gamma$ just beyond the bifurcation point. However, the simulations in \citet{Helfrich1995} did not impose a constant pressure-drop condition, unlike the calculations in \citet{Wylie1995} and the present work. Instead, \citet{Helfrich1995} required the laterally-averaged, mean flow rate to remain constant, with an inlet condition of either a laterally-uniform flow rate $q_z$ or a laterally-uniform inlet pressure $P_{in}$. Further study of the nonlinear evolution of thermoviscous channels is clearly needed, especially with regard to the role of boundary conditions, as well as cross-gap variations. The WRIBL model will allow for the efficient exploration of both issues.

\subsection{Linearly decreasing wall temperature}\label{sec:Slow}

In the previous section, we noted that both averaged models performed poorly near the entrance of the channel where the temperature of the hot fluid reduces rapidly on encountering the cold walls (cf. figure~\ref{fig:basetheta}). Indeed, this zone near the inlet violates the basic prerequisite for the validity of any long-wave, thin-gap averaged model, namely that of slow in-plane variations. As such, the case of a constant wall-temperature is a particularly stringent test for the Darcy and WRIBL models. A better-suited scenario for long-wave averaging would be one where the temperature of the walls starts out the same as the hot fluid and then gradually decreases along the flow direction. Would the WRIBL model provide a significant improvement over the Darcy model even in such a case, or are cross-gap variations not as significant? With this question in mind, we now consider a wall temperature that varies linearly along the flow direction from $T_H$ to $T_C$, or in non-dimensional terms, $T_w = 1-z$.

\subsubsection{Uniform base state flow}

\begin{figure}
\centering
	\subfloat[]{\includegraphics[width=0.49\textwidth]{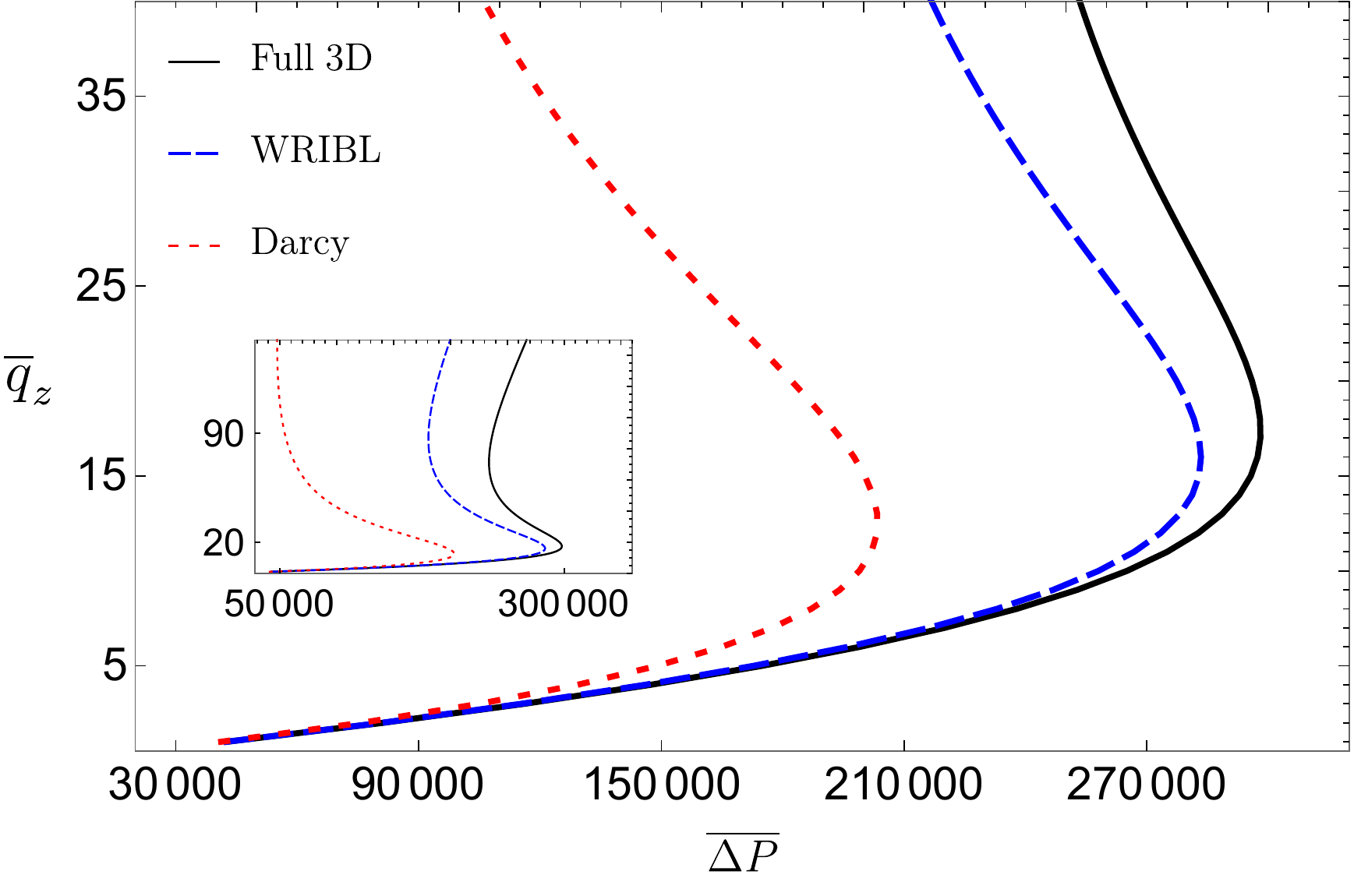}\label{fig:lintemp_base}}\hspace{1pt}
    \subfloat[]{\includegraphics[width=0.49\textwidth]{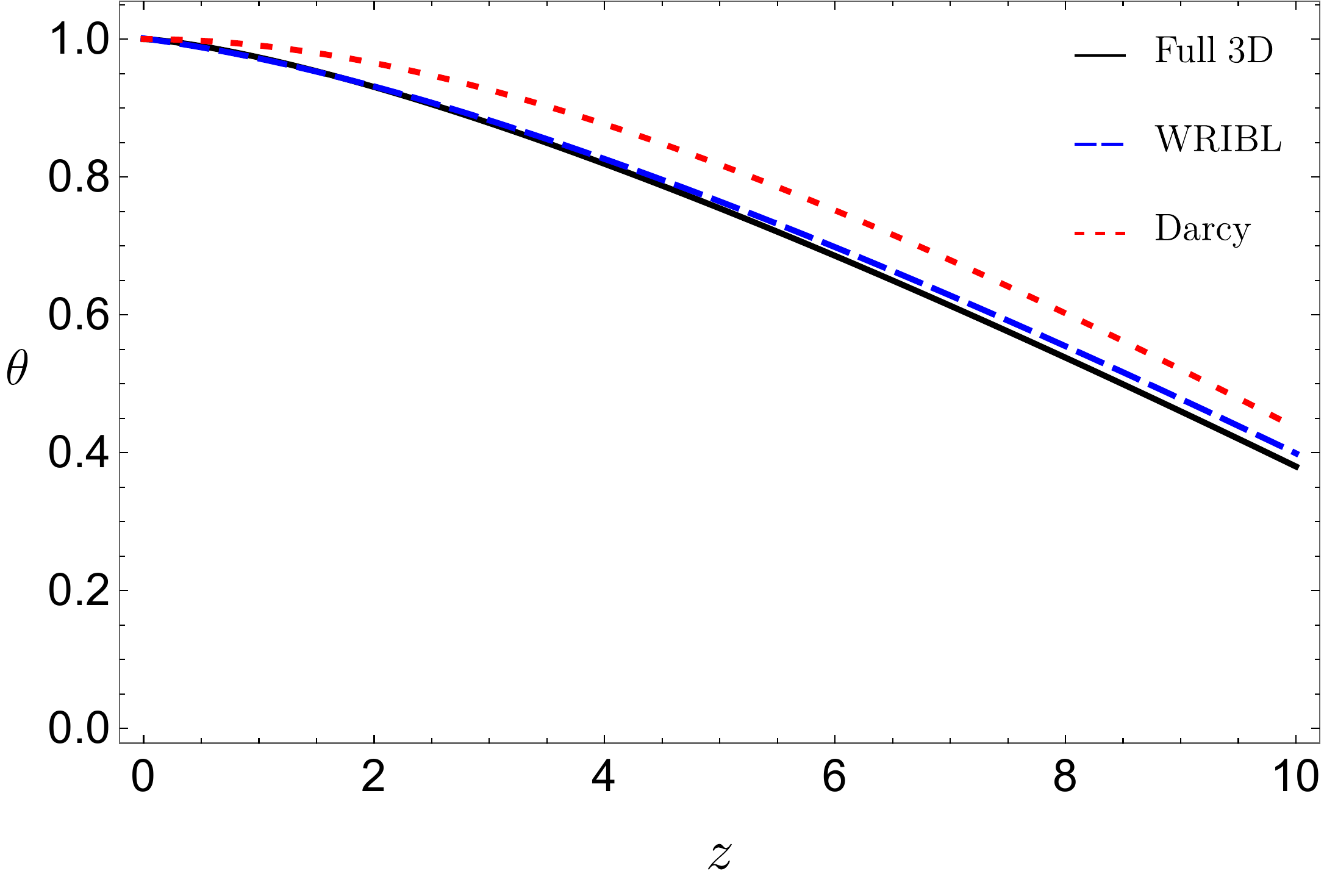}\label{fig:lintemp_basetheta}}
     \caption{
     (a) The base state flow rate ($\overline{q}_z$) as a function of the applied pressure drop ($\overline{\Delta P}$), as predicted by all three models (see the legend), for the case of a linearly decreasing wall temperature. The inset shows the behaviour over a wide range of $\overline{q}_z$, while the main panel focuses on lower values of $\overline{q}_z$. (b) Evolution of the gap-averaged base state temperature ($\overline{\theta}$) along the primary flow direction ($z$), for the state with $\overline{q}_z$=50. Parameter values: $L=10$ and $\gamma=8$.}
\end{figure}

We begin with a comparison of the uniform base states, specifically of the flow rate -- pressure drop relationship, which is presented in Figure~\ref{fig:lintemp_base} for the 3D (solid-black), WRIBL (blue-dashed) and Darcy (red-dashed) models. The inset shows the full S-shaped curves while the main panel zooms into the low-flow-rate regime which is of primary interest. Clearly, the WRIBL model is much more accurate than the Darcy model, even more so than in the case of the constant wall-temperature (cf. figure~\ref{fig:base}). Turning to the evolution of the gap-averaged temperature, depicted in Figure~\ref{fig:lintemp_basetheta}, we see that the WRIBL results nearly overlap with those of the 3D model, while the Darcy model shows a small offset. So, even though the Darcy model predicts the gap-averaged temperature reasonably well, its neglect of cross-gap variations in temperature and therefore in viscosity result in large errors in the prediction of the flow rate (figure~\ref{fig:lintemp_base}).

\begin{figure}
\centering
	\subfloat[]{\includegraphics[width=0.33\textwidth]{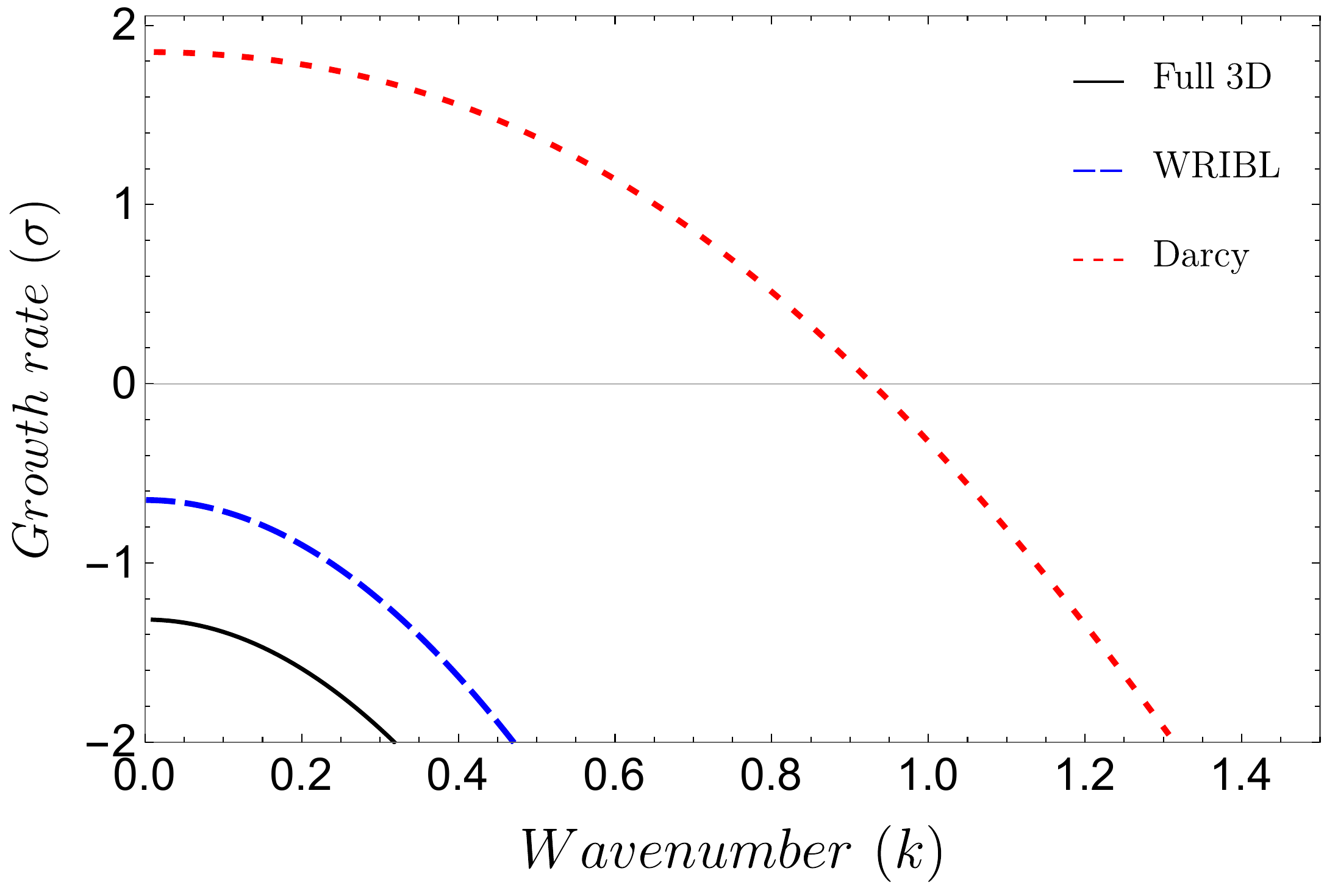}\label{fig:lintemplsaq15}}\hspace{1pt}
     \subfloat[]{\includegraphics[width=0.33\textwidth]{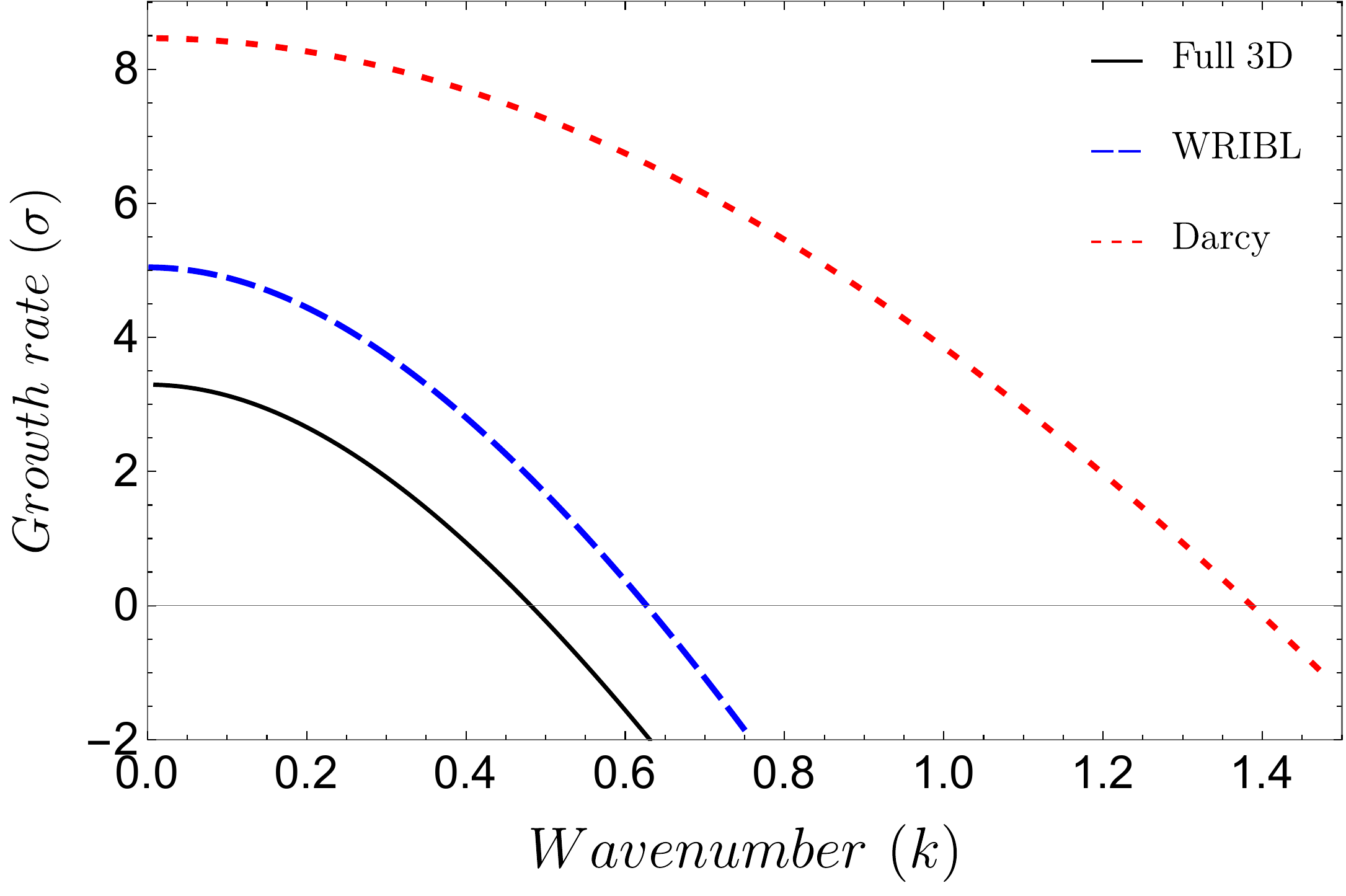}\label{fig:lintemplsaq25}}\hspace{1pt}
     \subfloat[]{\includegraphics[width=0.33\textwidth]{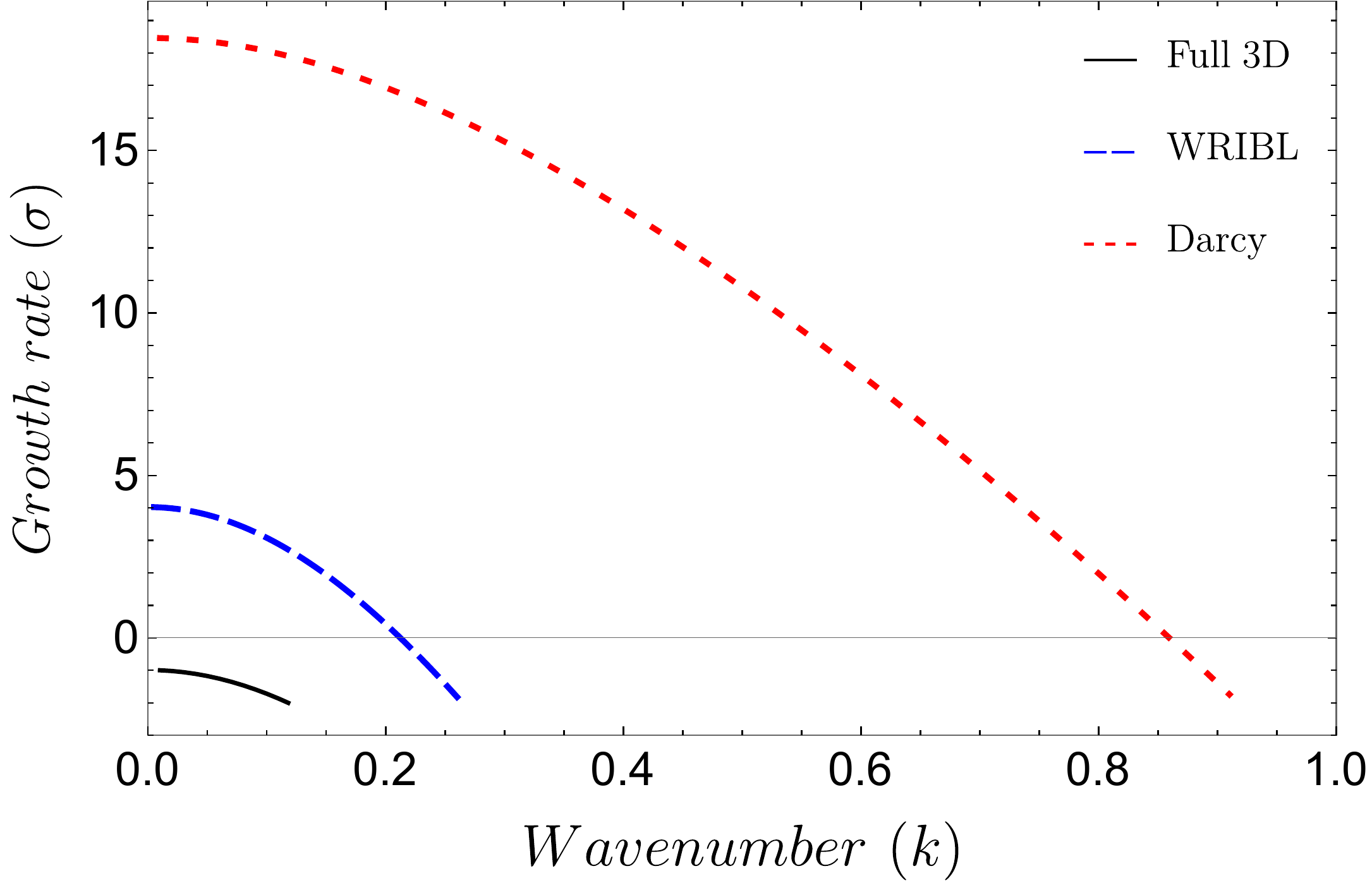}\label{fig:lintemplsaq75}}
     \caption{
     Comparison of the linear stability results of the three models (see the legend) for the case of a linearly decreasing wall temperature. The growth rate is plotted as a function of the wavenumber of the perturbation, for three different base states corresponding to flow rates of (a) $\bar{q}_z=15$, (b) $\bar{q}_z=25$, (c) $\bar{q}_z=75$. Parameter values: $L=10$, $\gamma=8$.}\label{fig:lintemplsa}
\end{figure}

\subsubsection{Linear instability}

Next, we consider the linear stability predictions of the three models, which are shown in figure~\ref{fig:lintemplsa} for base states corresponding to three flow rates: (a) $\bar{q}_z=15$, (b) $\bar{q}_z=25$ and (c) $\bar{q}_z=75$. We find that, for all three models, the base state is unstable only along the backward branch and the highest growth rate always corresponds to $k=0$. This is reflected in the three panels of figure~\ref{fig:lintemplsa}. The flow rates chosen correspond to values that lie on the lower, backward and upper branches of the flow-rate -- pressure-drop curve of the 3D model. Thus the growth curves of the 3D model (black-solid) show instability only for the intermediate case in figure~\ref{fig:lintemplsaq25}. The growth rates of the WRIBL model (blue-dashed) are quite close to those of the 3D model, whereas the Darcy model predicts much higher values (red-dotted). 

Qualitatively, though, all the growth-rate curves display similar features: a maximum at $k=0$ and a diffusive-like stabilization of higher wavenumbers. Such curves typically do not give rise to well-ordered patterns~\citep{Cross}, although irregular, long-length-scale, lateral modulations are possible. Most likely, though, the system will eventually evolve to a slow or fast uniform flow state corresponding to the stable lower or upper branch, respectively. However, a conclusive statement requires nonlinear simulations which we leave for a future study.

Comparing figures~\ref{fig:lsa1} and~\ref{fig:lintemplsa}, we see that the gradual variation of the wall temperature allows for a much better description of the system's stability by averaged models, provided the cross-gap variations are taken into account, as is done in the WRIBL model. 

\section{Summary and concluding remarks}\label{sec:summary}
A thin-gap averaged nonlinear model is presented, using the weighted residual integral boundary layer (WRIBL) method, for flow of a fluid with non-uniform viscosity. The model is applied to the problem of thermoviscous fingering in a Hele-Shaw geometry, wherein the dependence of viscosity on temperature leads to the spontaneous formation of alternate bands of low-viscosity (fast flowing) hot fluid and high viscosity (slow flowing) cold fluid. The cross-gap variation in temperature, which is crucial to the physics of the problem, results in a strongly non-parabolic flow profile across the thin gap. The WRIBL model accounts for these cross-gap variations, which were neglected by the \textit{ad hoc} Darcy model used in previous work. Comparing the predictions of the two averaged models with fully three-dimensional calculations we find that, although the results are in qualitative agreement, the WRIBL model performs much better than the Darcy model. Therefore, when quantitative accuracy is important, e.g., for comparison with experiments, the WRIBL model ought to be used. Furthermore, the WRIBL model will be crucial when studying issues that depend critically on the shape of the cross-gap velocity profile, such as laminar dispersion of solutes~\citep{Dutta2009} and inertial focusing of particles~\citep{Carlo2007}.

The broader implications of this work are twofold. 
First, it is shown that the averaging of the variable-viscosity transport equation (WRIBL model) is not the same as introducing variable viscosity into an averaged equation (Darcy model). Second, the ideas behind this work go beyond magma flow dynamics in fissures to other thin-gap problems, such as the Hele-Shaw flow of solutions or suspensions, where the viscosity varies due to gradients in solute or particle concentration. In such cases, the energy conservation equation for temperature would have to be replaced by an appropriate transport equation for the solute/particle concentration, which may then be coupled to the WRIBL averaged flow model by specifying the dependence of viscosity on concentration. 
Moreover, the WRIBL procedure presented here can also be used to develop reduced-order evolution equations for thin film interfacial flows, in which the viscosity varies across the depth of the thin film, e.g., the flow of a thin film of a particle suspension, in which the particles preferentially accumulate at the free-surface due to shear-induced migration~\citep{Morris2005}. Indeed, the WRIBL method was developed for thin film problems and easily accounts for a deforming free surface. 

Finally, it is important to note that the primary contribution of this work is the WRIBL averaging of the \textit{momentum equations} with a variable viscosity, not the averaging of the thermal transport equation. The latter was done as part of the illustrative example of thermoviscous fingering, in order to compare the WRIBL model with the Darcy-like model of \citet{Helfrich1995}, wherein both momentum and energy equations were averaged. In general, however, one need not average the transport equation for the scalar that determines the viscosity. Indeed, one could conceive of an iterative/time-stepping solution scheme in which the full transport equation is solved along with the WRIBL averaged flow model. As the transport equation is linear, such an approach would still be far more efficient, computationally, than solving the full set of nonlinear momentum and transport equations. This ``partial-averaging" strategy would also be more accurate than a fully averaged model, particularly in regimes where convective transport is important.

\vspace{\baselineskip}
D.S.P. and R.N. acknowledge financial support from NASA (NNX17AL27G) and NSF (2025117). R.N. is grateful to the Institute of Advanced Study-Durham University, UK for a fellowship and to Prof. E.W. Llewellin for introducing him to the physics of magma flows. J.R.P. is grateful for funding from the IIT Bombay IRCC Seed Grant. D.S.P. acknowledges support from the IIT Kanpur Initiation Grant, and DST-SERB grant SRG/2020/00242.

\bibliographystyle{jfm}
\bibliography{JFM-Volcano}

\begin{thebibliography}{37}
\expandafter\ifx\csname natexlab\endcsname\relax\def\natexlab#1{#1}\fi
\def\au#1{#1} \def\ed#1{#1} \def\yr#1{#1}\def\at#1{#1}\def\jt#1{\textit{#1}}
  \def\bt#1{#1}\def\bvol#1{\textbf{#1}} \def\vol#1{#1} \def\pg#1{#1}
  \def\publ#1{#1}\def\arxiv#1{#1}\def\org#1{#1}\def\st#1{\textit{#1}}

\bibitem[Balakotaiah \& Ratnakar(2010)]{Balakotaiah2010}
{\sc \au{Balakotaiah, Vemuri} \& \au{Ratnakar, Ram~R.}} \yr{2010}  \at{On the
  use of transfer and dispersion coefficient concepts in low-dimensional
  diffusion–convection-reaction models}.  \jt{Chem. Eng. Res. Des.}
  \bvol{88}~(3),  \pg{342--361}.

\bibitem[Bercovici(1994)]{Bercovici1994}
{\sc \au{Bercovici, D.}} \yr{1994}  \at{A theroretical model of cooling viscous
  gravity currents with temperature-dependent viscosity}.  \jt{Geophysical
  Research Letters}  \bvol{21},  \pg{1177--1180}.

\bibitem[Buongiorno(2005)]{Buongiorno2005}
{\sc \au{Buongiorno, J.}} \yr{2005}  \at{Convective transport in nanofluids}.
  \jt{J. Heat Transf.}  \bvol{128}~(3),  \pg{240--250}.

\bibitem[Carlo {\em et~al.\/}(2007)Carlo, Irimia, Tompkins \& Toner]{Carlo2007}
{\sc \au{Carlo, D.~Di}, \au{Irimia, D.}, \au{Tompkins, R.~G.} \& \au{Toner,
  M.}} \yr{2007}  \at{Continuous inertial focusing, ordering, and separation of
  particles in microchannels}.  \jt{Proc Natl Acad Sci USA}  \bvol{104}~(48),
  \pg{18892--18897}.

\bibitem[Choudhary {\em et~al.\/}(2020)Choudhary, Li, Renganathan, Xuan \&
  Pushpavanam]{choudhary2020}
{\sc \au{Choudhary, A.}, \au{Li, D.}, \au{Renganathan, T.}, \au{Xuan, X.} \&
  \au{Pushpavanam, S.}} \yr{2020}  \at{Electrokinetically enhanced cross-stream
  particle migration in viscoelastic flows}.  \jt{J. Fluid Mech.}  \bvol{898},
  \pg{A20}.

\bibitem[Cross \& Greenside(2009)]{Cross}
{\sc \au{Cross, M.} \& \au{Greenside, H.}} \yr{2009} {\em Pattern Formation and
  Dynamics in Nonequilibrium Systems\/}.  \publ{Cambridge University Press}.

\bibitem[Datta \& Ghosal(2009)]{Dutta2009}
{\sc \au{Datta, S.} \& \au{Ghosal, S.}} \yr{2009}  \at{Characterizing
  dispersion in microfluidic channels}.  \jt{Lab Chip}  \bvol{9},
  \pg{2537--2550}.

\bibitem[Dietze \& Ruyer-Quil(2015)]{Dietze2015}
{\sc \au{Dietze, G.} \& \au{Ruyer-Quil, C.}} \yr{2015}  \at{Film in narrow
  tubes}.  \jt{J.~Fluid Mech.}  \bvol{762},  \pg{68–109}.

\bibitem[Dietze \& Ruyer-Quil(2013)]{Dietze2013}
{\sc \au{Dietze, G.~F.} \& \au{Ruyer-Quil, C.}} \yr{2013}  \at{Wavy liquid
  films in interaction with a confined laminar gas flow.}  \jt{J.~Fluid Mech.}
  \bvol{722},  \pg{348--393}.

\bibitem[Helfrich(1995)]{Helfrich1995}
{\sc \au{Helfrich, K.~R.}} \yr{1995}  \at{Thermo-viscous fingering of flow in a
  thin gap: a model of magma flow in dikes and fissures}.  \jt{J.~Fluid Mech.}
  \bvol{305},  \pg{219--238}.

\bibitem[Homsy(1987)]{Homsy1987}
{\sc \au{Homsy, G.~M.}} \yr{1987}  \at{Viscous fingering in porous media}.
  \jt{Ann. Rev. Fluid Mech.}  \bvol{19},  \pg{271--311}.

\bibitem[Huppert(2002)]{Huppert2002}
{\sc \au{Huppert, H.}} \yr{2002}  \at{{Geological fluid mechanics}}.  \bt{In
  {\em In Perspectives in Fluid Dynamics\/} (ed. \ed{G.~K. Batchelor, H.~K.
  Moffatt \& M.~G. Worster})},  \pg{p. 447}.  \publ{Cambridge University
  Press}.

\bibitem[Kalliadasis {\em et~al.\/}(2012)Kalliadasis, Scheid, Ruyer-Quil \&
  Velarde]{Kalliadasis}
{\sc \au{Kalliadasis, S.}, \au{Scheid, B.}, \au{Ruyer-Quil, C.} \& \au{Velarde,
  M.~G.}} \yr{2012} {\em Falling Liquid Films\/}.  \publ{Springer}.

\bibitem[Leighton \& Acrivos(1987)]{Acrivos1987}
{\sc \au{Leighton, D.} \& \au{Acrivos, A.}} \yr{1987}  \at{The shear-induced
  migration of particles in concentrated suspensions}.  \jt{J.~Fluid Mech.}
  \bvol{181},  \pg{415--439}.

\bibitem[Lyon \& Leal(1998)]{Lyon1998}
{\sc \au{Lyon, M.~K.} \& \au{Leal, L.~G.}} \yr{1998}  \at{An experimental study
  of the motion of concentrated suspensions in two-dimensional channel flow.
  part 1. monodisperse systems}.  \jt{J.~Fluid Mech.}  \bvol{363},
  \pg{25--56}.

\bibitem[Marmet {\em et~al.\/}(2017)Marmet, Scacchi \& Brader]{Marmet2017}
{\sc \au{Marmet, P.}, \au{Scacchi, A.} \& \au{Brader, J.~M.}} \yr{2017}
  \at{Shear-induced migration in colloidal suspensions}.  \jt{Mol.~Phys.}
  \bvol{115}~(14),  \pg{1691--1699}.

\bibitem[Nagatsu {\em et~al.\/}(2009)Nagatsu, Fujita, Kato \&
  Tada]{Nagatsu2009}
{\sc \au{Nagatsu, Y.}, \au{Fujita, N.}, \au{Kato, Y.} \& \au{Tada, Y.}}
  \yr{2009}  \at{An experimental study of non-isothermal miscible displacements
  in a hele-shaw cell}.  \jt{Experimental Thermal and Fluid Science}
  \bvol{33},  \pg{695--705}.

\bibitem[Nott \& Brady(1994)]{Nott1994}
{\sc \au{Nott, P.~R.} \& \au{Brady, J.~F.}} \yr{1994}  \at{Pressure-driven flow
  of suspensions: simulation and theory}.  \jt{J.~Fluid Mech.}  \bvol{275},
  \pg{157--199}.

\bibitem[Oron \& Heining(2008)]{Oron2008}
{\sc \au{Oron, A.} \& \au{Heining, C.}} \yr{2008}  \at{Weighted-residual
  integral boundary-layer model for the nonlinear dynamics of thin liquid films
  falling on an undulating vertical wall}.  \jt{Geophysical Research Letters}
  \bvol{20},  \pg{082102}.

\bibitem[Pearson {\em et~al.\/}(1973)Pearson, Shah \& Vieira]{Pearson1973}
{\sc \au{Pearson, J. R.~A.}, \au{Shah, Y.~T.} \& \au{Vieira, E. S.~A.}}
  \yr{1973}  \at{Stability of non-isothermal flow in channels -- {I}.
  {T}emperature dependent {N}ewtonian fluid without heat generation}.
  \jt{Chemical Engineering Science}  \bvol{28},  \pg{2079--2088}.

\bibitem[Pradas {\em et~al.\/}(2014)Pradas, Tseluiko, Ruyer-Quil \&
  Kalliadasis]{Pradas2014}
{\sc \au{Pradas, M.}, \au{Tseluiko, D.}, \au{Ruyer-Quil, C.} \&
  \au{Kalliadasis, S.}} \yr{2014}  \at{Pulse dynamics in a power-law falling
  film}.  \jt{J.~Fluid Mech.}  \bvol{747},  \pg{460--480}.

\bibitem[Pritchard(2009)]{Pritchard2009}
{\sc \au{Pritchard, D.}} \yr{2009}  \at{The linear stability of
  double-diffusive miscible rectilinear displacements in a hele-shaw cell}.
  \jt{European Journal of Mechanics - B/Fluids}  \bvol{28},  \pg{564--577}.

\bibitem[Roberts(2001)]{Roberts2001}
{\sc \au{Roberts, A.J}} \yr{2001}  \at{Holistic projection of initial
  conditions onto a finite difference approximation}.  \jt{Comput. Phys.
  Commun.}  \bvol{142}~(1),  \pg{316--321}.

\bibitem[Roberts(1992)]{Roberts1992}
{\sc \au{Roberts, A.~J.}} \yr{1992}  \at{Boundary conditions for approximate
  differential equations}.  \jt{J. Aust. Math. Soc. B}  \bvol{34}~(1),
  \pg{54–80}.

\bibitem[Roberts(2015)]{Robertsbook}
{\sc \au{Roberts, A.~J.}} \yr{2015} {\em Model Emergent Dynamics in Complex
  Systems\/}.  \publ{SIAM}.

\bibitem[Roberts \& Bunder(2017)]{Roberts2017}
{\sc \au{Roberts, A.~J.} \& \au{Bunder, J.~E.}} \yr{2017}  \at{{Slowly varying,
  macroscale models emerge from microscale dynamics over multiscale domains}}.
  \jt{IMA J. Appl. Math.}  \bvol{82}~(5),  \pg{971--1012}.

\bibitem[Ruyer-Quil(2001)]{RQ2001}
{\sc \au{Ruyer-Quil, C.}} \yr{2001}  \at{Inertial corrections to the darcy law
  in a hele–shaw cell}.  \jt{C. R. Acad. Sci. Paris}  \bvol{329},  \pg{1--6}.

\bibitem[Ruyer-Quil {\em et~al.\/}(2012)Ruyer-Quil, Chakraborty \&
  Dandapat]{RQ2012}
{\sc \au{Ruyer-Quil, C.}, \au{Chakraborty, S.} \& \au{Dandapat, B.~S.}}
  \yr{2012}  \at{Wavy regime of a power-law film flow}.  \jt{J.~Fluid Mech.}
  \bvol{692},  \pg{220--256}.

\bibitem[Ruyer-Quil \& Manneville(2002)]{RQ2002}
{\sc \au{Ruyer-Quil, C.} \& \au{Manneville, P.}} \yr{2002}  \at{Further
  accuracy and convergence results on the modeling of flows down inclined
  planes by weighted-residual approximations}.  \jt{Phys. Fluids}  \bvol{14},
  \pg{170--183}.

\bibitem[Timberlake \& Morris(2005)]{Morris2005}
{\sc \au{Timberlake, B.~D.} \& \au{Morris, J.~F.}} \yr{2005}  \at{Particle
  migration and free-surface topography in inclined plane flow of a
  suspension}.  \jt{J.~Fluid Mech.}  \bvol{538},  \pg{309--341}.

\bibitem[Trevelyan \& Kalliadasis(2004)]{Trev2004}
{\sc \au{Trevelyan, P. M.~J.} \& \au{Kalliadasis, S.}} \yr{2004}  \at{Wave
  dynamics on a thin-film falling down a heated wall.}  \jt{J.~Engg. Math.}
  \bvol{50},  \pg{177--208}.

\bibitem[Vennamneni {\em et~al.\/}(2020)Vennamneni, Nambiar \&
  Subramanian]{Laxmi2020}
{\sc \au{Vennamneni, L.}, \au{Nambiar, S.} \& \au{Subramanian, G.}} \yr{2020}
  \at{Shear-induced migration of microswimmers in pressure-driven channel
  flow}.  \jt{Journal of Fluid Mechanics}  \bvol{890}.

\bibitem[Viola {\em et~al.\/}(2017)Viola, Gallaire \& Dollet]{Viola2017}
{\sc \au{Viola, F.}, \au{Gallaire, F.} \& \au{Dollet, B.}} \yr{2017}
  \at{Sloshing in a {H}ele-{S}haw cell: experiments and theory}.  \jt{J.~Fluid
  Mech.}  \bvol{831},  \pg{R1}.

\bibitem[Whitehead \& Griffiths(2001)]{Balm2001}
{\sc \au{Whitehead, J.~A.} \& \au{Griffiths, R.~W.}} \yr{2001}
  \at{{Morphological Instabilities in Flows with Cooling, Freezing or
  Dissolution}}.  \bt{In {\em In Geomorphological Fluid Mechanics\/} (ed.
  \ed{N.~J. Balmforth \& A.~Provenzale})},  \pg{pp. 139--144}.
  \publ{Springer}.

\bibitem[Whitehead \& Helfrich(1991)]{Whitehead1991}
{\sc \au{Whitehead, J.~A.} \& \au{Helfrich, K.~R.}} \yr{1991}  \at{Instability
  of flow with temperature-dependent viscosity: {A} model of magma dynamics}.
  \jt{J. Geophysical Research}  \bvol{96},  \pg{4145--4155}.

\bibitem[Wylie {\em et~al.\/}(1999)Wylie, Helfrich, Dade, Lister \&
  Salzig]{Wylie1999}
{\sc \au{Wylie, J.~J.}, \au{Helfrich, K.~R.}, \au{Dade, B.}, \au{Lister, J.~R.}
  \& \au{Salzig, J.~F.}} \yr{1999}  \at{Flow localization in fissure
  eruptions}.  \jt{Bull. Volcanol.}  \bvol{60},  \pg{432--440}.

\bibitem[Wylie \& Lister(1995)]{Wylie1995}
{\sc \au{Wylie, J.~J.} \& \au{Lister, J.~R.}} \yr{1995}  \at{The effects of
  temperature-dependent viscosity on flow in a cooled channel with application
  to basaltic fissure eruptions}.  \jt{J.~Fluid Mech.}  \bvol{305},
  \pg{239--261}.

\end{thebibliography}

\end{document}